# Incorporation of Polar Mellin Transform in a Hybrid Optoelectronic Correlator for Scale and Rotation Invariant Target Recognition

**Mehjabin Sultana Monjur,[1,*] Shih Tseng,[1,4] Renu Tripathi,[3] and M.S. Shahriar[1,2]**

[1]*Department of Electrical Engineering and Computer Science, Northwestern University, Evanston, IL 60208, USA*

[2]*Department of Physics and Astronomy, Northwestern University, Evanston, IL 60208, USA*

[3]*OSCAR, Department of Physics and Engineering, Delaware State University, Dover, DE 19901, USA*

[4]*Digital Optics Technologies, Rolling Meadows, IL 60008, USA*

[*]*Corresponding author: mehjabin@u.northwestern.edu*

In this paper, we show that our proposed hybrid optoelectronic correlator (HOC), which correlates images using spatial light modulators (SLM), detectors and field programmable gate arrays (FPGAs), is capable of detecting objects in a scale and rotation invariant manner, along with the shift invariance feature, by incorporating polar mellin transform (PMT). We also illustrate a key limitation of the ideas presented in previous papers on performing PMT and present a solution to circumvent this limitation by cutting out a small circle at the center of the Fourier Transform which precedes PMT. Furthermore, we show how to carry out shift, rotation and scale invariant detection of multiple matching objects simultaneously, a process previously thought to be incompatible with PMT based correlators. We present results of numerical simulations to validate the concepts.

*OCIS codes:* (070.0070) Fourier optics and signal processing; (100.0100) Image Processing; (130.0130) Integrated Optics; (070.4550) Correlators; (100.3005) Image recognition devices; (130.0250) Optoelectronics

## 1. Introduction

Target identification and tracking is important in many defense and civilian applications. Optical correlators provide a simple technique for fast verification and identification of data. The simplest form of such a device is the basic Vander Lugt [1-3] optical correlator .The limitation of this correlator is that the recording process is very time consuming. This constraint is circumvented in a joint transform correlator (JTC) [4-6], where a dynamic material such as photorefractive polymer film is used so that the recording and correlation take place simultaneously. A JTC of this type suffers from many practical problems and constraints. As an alternative, we recently proposed a hybrid optoelectronic correlator (HOC) [7] where the non-linearity provided by the JTC medium is replaced by the non-linearity of high-speed detectors. The advantage of this approach as compared to previous architecture that also employ detectors [15] has been discussed in details in ref. 7. As shown in ref [7], the HOC is capable of shift-invariant image recognition, in the same manner as what is achieved with a conventional holographic correlator (CHC). However, the actual architecture is very different, requiring many intermediate steps, servos and post processing. Furthermore, the output signals are also different (e.g., it contains a cross-correlation term and a reverse cross-correlation term, but no convolution nor dc term).Thus, it may not be a priori obvious whether the HOC can also perform scale and rotation invariant correlation using polar mellin transform (PMT) [8-13].

In this paper, we show how, indeed, it is possible to use PMT to achieve scale and rotation invariant image recognition with the HOC architecture. Furthermore, we identify some limitations of using PMT in the CHC architecture proposed previously, and show how to overcome these constraints by proper pre-processing of images for both CHC and HOC.

The paper is organized as follows: Section 2 describes the proposed HOC architecture incorporating PMT to achieve scale and rotation invariance in addition to shift invariance. Section 3 presents a brief review of the underlying concepts of PMT for achieving scale and rotation invariance. Section 4 illustrates the simulation results using MATLAB, which shows that the proposed system is capable of detecting an image in a scale and rotation invariant manner. In this section, we also point out some limitations of using PMT in the CHC architecture as proposed previously, and show how to overcome these constraints. Section 5 describes the process of detecting multiple objects in a shift, scale and rotation invariant manner using the HOC architecture and show the simulation results. The paper concludes with a summary and outlook in section 6.

## 2. Proposed Hybrid Optoelectronic Correlator incorporating PMT

The details of the HOC proposed by us can be found in reference 7. Briefly, in the shift invariant HOC architecture, the reference image is interfered with a plane wave and the resulting signal is recorded with a detector array. Similarly, the object image is also interfered with a plane wave, and the resulting signal is recorded with another detector array. These signals are then processed in a certain manner, as described below, and then sent to an SLM to perform the correlation in the optical domain.

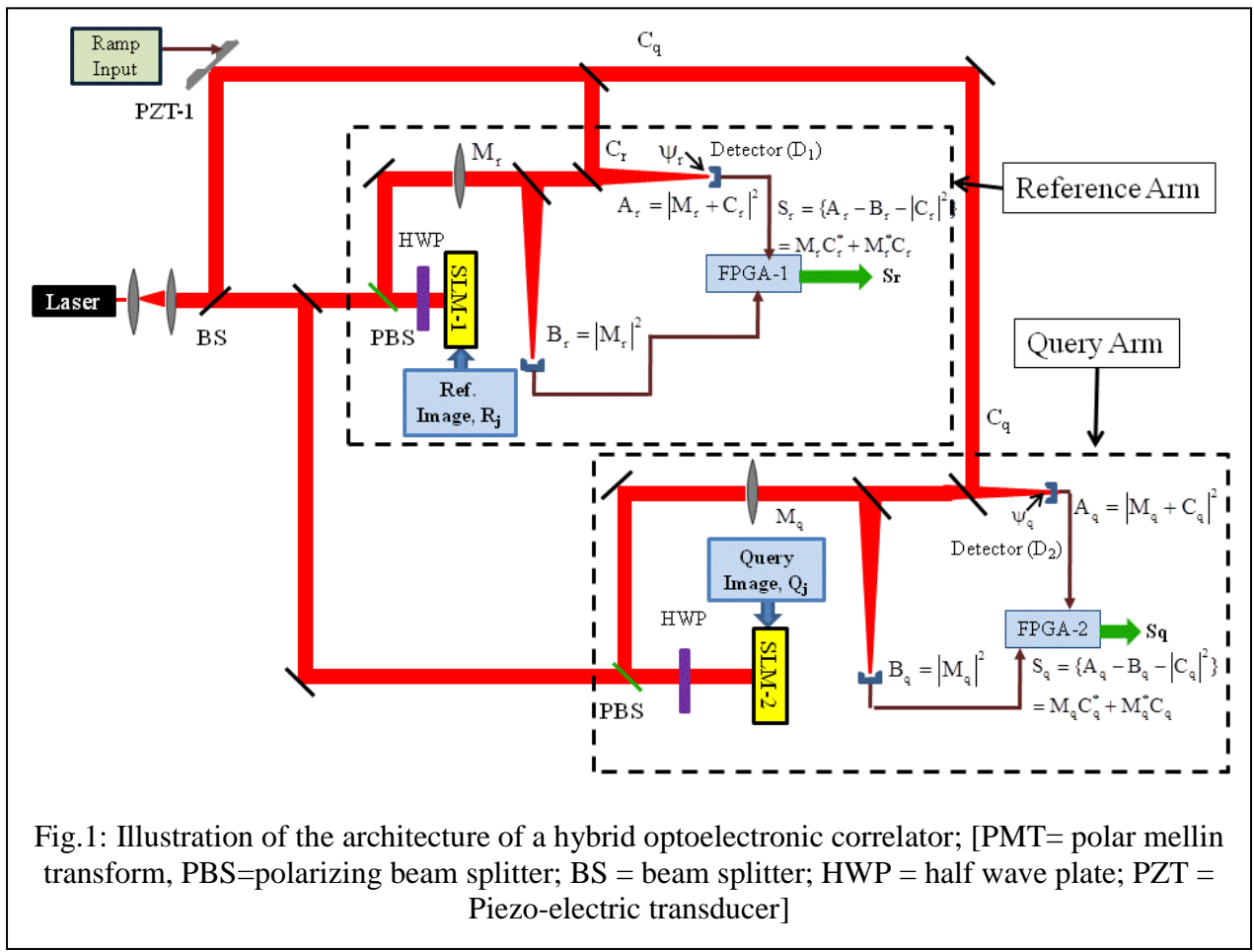

Fig.1: Illustration of the architecture of a hybrid optoelectronic correlator; [PMT= polar mellin transform, PBS=polarizing beam splitter; BS = beam splitter; HWP = half wave plate; PZT = Piezo-electric transducer]

Fig. 1 shows an overview of the HOC architecture, which can be used to correlate two images, $R_j$ and $Q_j$. Here, $R_j$ and $Q_j$, can each be a conventional image, or the PMT processed version of a conventional image. The process for producing a PMT processed version of a conventional image will be described later on in this section. We envision a scenario where a set of reference images, $R_j$, are stored in a database. A particular image of interest is then retrieved from the database for correlation. If a digital database (such as computer hard drive) is used, then the image is converted to the optical domain using an SLM. The optical image ($R_j$) is Fourier Transformed by a lens. It is then split into two identical ports, both designated as $M_r$. In one port, the image is detected by an array of detectors, which could be a high resolution focal plane array (FPA) or a digital CMOS camera. The signal array produced by the camera is denoted as $B_r$. The camera is interfaced with a field programmable gate array (FPGA) via a USB cable. $B_r$ can be stored in the built in memory of the FPGA [FPGA-1]. In the other port, $M_r$ is interfered with a plane wave $C_r$, and detected with another CMOS camera, producing the digital signal array $A_r$ and is stored in the memory of FPGA-1. $A_r$ and $B_r$ can be expressed as:

$$A_r = \left| M_r e^{j\phi_r} + C_r e^{j\psi_r} \right|^2 = |M_r|^2 + |C_r|^2 + |M_r||C_r|e^{j(\phi_r - \psi_r)} + |M_r||C_r|e^{-j(\phi_r - \psi_r)} \quad (1)$$

$$B_r = |M_r|^2 \quad (2)$$

In addition, the intensity profile of the plane wave ($|C_r|^2$) is measured, by blocking the image path momentarily, using a shutter (not shown), and the information is stored in the memory component of FPGA-1. FPGA-1 then computes and stores $S_r$ which can be expressed as:

$$S_r = A_r - B_r - |C_r|^2 = M_r C_r^* + M_r^* C_r = |M_r||C_r|e^{j(\phi_r - \psi_r)} + |M_r||C_r|e^{-j(\phi_r - \psi_r)} \quad (3)$$

Here $\phi_r$ is the phase of the Fourier transformed image, $M_r$, and $\Psi_r$ is the phase of the plane wave, $C_r$. It should be noted that $\phi$ is a function of $(x,y)$, assuming that the image is in the $(x,y)$ plane. This subtraction process has to be done pixel by pixel using one or more subtractors available in the FPGA.

The captured query image, $Q_j$ is also transferred to the optical domain using another SLM (SLM-2) to form $Q_j$. The optical image ($Q_j$) is Fourier Transformed by a lens and is split into two identical ports, both designated as $M_q$. In a manner similar to what is described above for the reference image, the signal $S_q = A_q - B_q - |C_q|^2$ is produced using two cameras and an FPGA (FPGA-2) and stored in FPGA-2 memory. Here, $C_q$ is the amplitude of an interfering plane wave, and the other quantities are given as follows:

$$A_q = |M_q + C_q|^2 = |M_q|^2 + |C_q|^2 + |M_q||C_q|e^{j(\phi_q-\psi_q)} + |M_q||C_q|e^{-j(\phi_q-\psi_q)} \quad (4)$$

$$B_q = |M_q|^2 \quad (5)$$

$$S_q = A_q - B_q - |C_q|^2 = M_q C_q^* + M_q^* C_q = |M_q||C_q|e^{j(\phi_q-\psi_q)} + |M_q||C_q|e^{-j(\phi_q-\psi_q)} \quad (6)$$

As before, $\phi_q(x,y)$ is the phase of the Fourier transformed image, $M_q$, and $\Psi_q$ is the phase of the plan wave $C_q$.

In the final stage of the hybrid correlator (as shown in fig 2), these two signals ($S_r$ and $S_q$) described in equations (3) and (6) are multiplied together using the multiplier in FPGA-3. Four quadrant multiplication can easily be implemented using an FPGA. The resulting signal array, S is stored in FPGA-3 memory. This can be expressed as:

$$S = S_r \cdot S_q = \left(M_r C_r^* + M_r^* C_r\right)\left(M_q C_q^* + M_q^* C_q\right)$$

$$= \left[\alpha^* M_r M_q + \alpha M_r^* M_q^* + \beta^* M_r M_q^* + \beta M_r^* M_q\right] \quad (7)$$

where, $\alpha \equiv |C_r||C_q|e^{j(\psi_r+\psi_q)}$; $\beta \equiv |C_r||C_q|e^{j(\psi_r-\psi_q)}$.

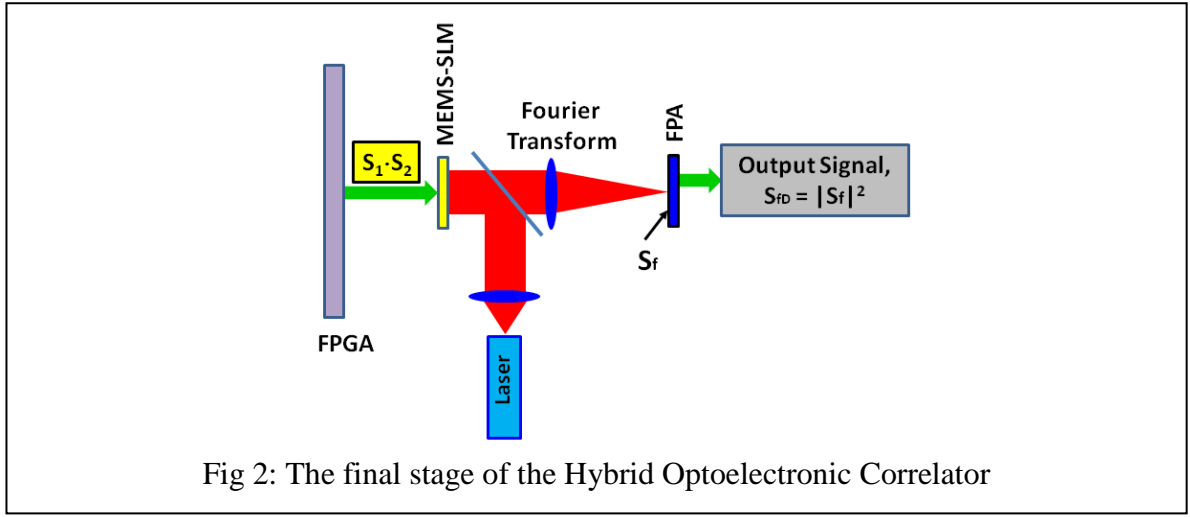

Fig 2: The final stage of the Hybrid Optoelectronic Correlator

This signal array, S, is now transferred to another SLM (SLM-3) from FPGA-3 through the Digital Visual Interface (DVI) port. Since S can be positive or negative, the SLM should be operated in a bipolar amplitude mode. The optical image produced by SLM-3 is Fourier transformed using a lens, producing the signal $S_f$, given by:

$$S_f = \left[\alpha^* F(M_r M_q) + \alpha F(M_r^* M_q^*) + \beta^* F(M_r M_q^*) + \beta F(M_r^* M_q)\right] \quad (8)$$

Here, F stands for the Fourier Transform. Since $M_\alpha$ ($\alpha$= r or q) is the Fourier Transform (FT) of the images $R_j$ or $Q_j$, using the well-known relations between the FT of products of functions, and convolutions and cross-correlations, we can express the final signal as the sum of four terms:

$$S_f = \alpha^* T_1 + \alpha T_2 + \beta^* T_3 + \beta T_4$$

$$\begin{aligned}
T_1 &= R_j(x,y) \otimes Q_j(x,y) \\
T_2 &= R_j^*(-x,-y) \otimes Q_j^*(-x,-y) \\
T_3 &= R_j(x,y) \odot Q_j(x,y) \\
T_4 &= Q_j(x,y) \odot R_j(x,y)
\end{aligned} \tag{9}$$

Where $\otimes$ indicates two-dimensional convolution, and $\odot$ indicates two-dimensional cross-correlation. We can now make the following observations:

- *$T_1$* represents the two-dimensional convolution of the images, $R_j$ and $Q_j$.
- *$T_2$* represents the two-dimensional convolution of the images, $R_j$ and $Q_j$, but with each conjugated and inverted along both axes.
- *$T_3$* represents the two-dimensional cross-correlation of the images, $R_j$ and $Q_j$.
- *$T_4$* represents the two-dimensional cross-correlation of the images, $Q_j$ and $R_j$. (Cross-correlation is non-commutative; hence, *$T_3$* is not necessarily equal to *$T_4$*). We denote it as anti-crosscorrelation signal.

The cross-correlation technique is usually used to find matches between two objects. In our architecture, we have convolution terms ($T_1$ and $T_2$) in addition to cross-correlation terms ($T_3$ and $T_4$). The convolution terms can be washed out by implementing a phase stabilization and scanning technique in the HOC architecture which has been discussed in details in ref [7].

The conventional holographic correlator (CHC) [2-4] has one convolution term and one cross-correlation term, along with some dc and low spatial frequency outputs. In contrast, the HOC architecture, after incorporating the phase stabilization and scanning, has only two outputs, a cross correlation ($T_3$) and an anti cross-correlation term ($T_4$). The strengths of $T_3$ and $T_4$ are the same, and they appear as symmetric signals in the presence of a shift. This difference has to be kept in mind in using the HOC architecture.

The signal observed by the final FPA, is, of course, given by $S_{fD} \equiv \alpha |S_f|^2$, where $\alpha$ is a proportionality constant, which we set to be unity for simplicity of discussion. Thus, assuming that the contributions from $T_1$ and $T_2$ are eliminated via the combination of phase scanning and low pass filtering [7], the final FPA signal can be expressed as:

$$S_{fD} = |\beta^* T_3 + \beta T_4|^2 = |\beta^* F(M_r M_q^*) + \beta F(M_r^* M_q)|^2 \tag{10}$$

It is useful to consider two different scenarios in order to interpret the information one can glean from this signal.

- *Scenario 1: Perfectly matched images with no relative shift:*

In this case, $T_3 = T_4$, so that eqn 10 can be expressed as: $S_{fD} = |\beta^* + \beta|^2 |T_3|^2 = 4|\beta|^2 \cos^2(\Psi_r - \Psi_q)|T_3|^2$. This signal is maximum when $(\Psi_r - \Psi_q) = 0$. As discussed in ref. 7, this situation (i.e. use of identical, unshifted images) thus can be used to keep the servo locked to the position where $(\Psi_r - \Psi_q) = 0$. Thus, we will assume from now on that $(\Psi_r - \Psi_q) = 0$, so that $\beta = \beta^* = |\beta|$.

- *Scenario 2: Perfectly matched images with a relative shift in position:*

If the query image is shifted by a vector, $\vec{\rho_0}$, then, $M_q = M_r \exp(i2\pi \vec{f} \cdot \vec{\rho_0})$. Hence, we can write: $M_r M_q^* = |M_r|^2 \exp(-i2\pi \vec{f} \cdot \vec{\rho_0})$; $M_r^* M_q = |M_r|^2 \exp(i2\pi \vec{f} \cdot \vec{\rho_0})$. Now we define: $F(|M_r|^2) \equiv G_0(\vec{\rho})$, where F stands for FT. It then follows that, $T_3 = F(M_r M_q^*) = F(|M_r|^2 \exp(-i2\pi \vec{f} \cdot \vec{\rho_0})) = G_0(\vec{\rho} - \vec{\rho_0})$. Similarly, $T_4 = F(M_r^* M_q) = F(|M_r|^2 \exp(i2\pi \vec{f} \cdot \vec{\rho_0})) = G_0(\vec{\rho} + \vec{\rho_0})$. The spatial extent of $G_0(\vec{\rho})$ is determined by the size of the image. Let us quantify this by defining a radial extent, $|\vec{\rho_m}|$ such that $G_0(\vec{\rho}) = 0$ for $|\vec{\rho}| \geq |\vec{\rho_m}|$. The behavior of the final signal depends on the value of the parameter, $\eta \equiv |\vec{\rho_0}|/|\vec{\rho_m}|$.

*Case I:* Consider first the situation where, $\eta \geq 1$. In this case, there is no overlap between $G_0(\vec{\rho} - \vec{\rho_0})$ and $G_0(\vec{\rho} + \vec{\rho_0})$ (i.e. between $T_3$ and $T_4$). Thus, the final detector signal of eqn 10 can be expressed as: $S_{fD} = |\beta|^2 (|T_3|^2 + |T_4|^2)$. In this case, we will see two distinct peaks, corresponding to the cross-correlation ($T_3$) and anti cross correlation ($T_4$). We would like to point out that, for the sake of simplicity, this condition was implicitly assumed to hold in the discussions presented in ref. 7.

*Case II:* Consider next the situation where, η< 1. In this case, the final detector signal can be expressed as: $S_{fD} = |\beta|^2|G_0(\vec{\rho} - \vec{\rho_0}) + G_0(\vec{\rho} + \vec{\rho_0})|^2 = |\beta|^2|T_3+T_4|^2$. The shape of this signal depends on the details of the images, and thereby on the details of $T_3$ and $T_4$. In what follows, we illustrate the shape of the signal $S_{fD}$ for both case I and II, with a few examples.

For clarity, we consider first examples of one-dimensional images. Figures 3(a) and 3(b) shows two images, with gaussian profiles, shifted from each other by $|\vec{\rho_0}| = 4\sigma$, where σ = 2.5 mm is the half-width of each image. Here, a reasonable estimate for $|\vec{\rho_m}|$ is 3σ, so that η > 1 is satisfied, corresponding to case I. As can be seen in fig 3(c), the signal now has two distinct peaks.

Next, we consider again the same images, but with a smaller shift: $|\vec{\rho_0}| = 2\sigma,$ shown in figures 3(d) and 3(e). In this case $\eta = 0.67$, corresponding to case II, so that there will be overlaps between $T_3$ and $T_4$. However, the two peaks can still be discerned in the final signal, shown in fig 3(f). Finally we consider a case where $|\vec{\rho_0}| = \sigma$, as shown in fig. 3(g) and 3(h). In this case, $\eta = 0.25$ and we can see that the peaks are no longer distinguishable, and have merged into each other, as shown in fig 3(i).

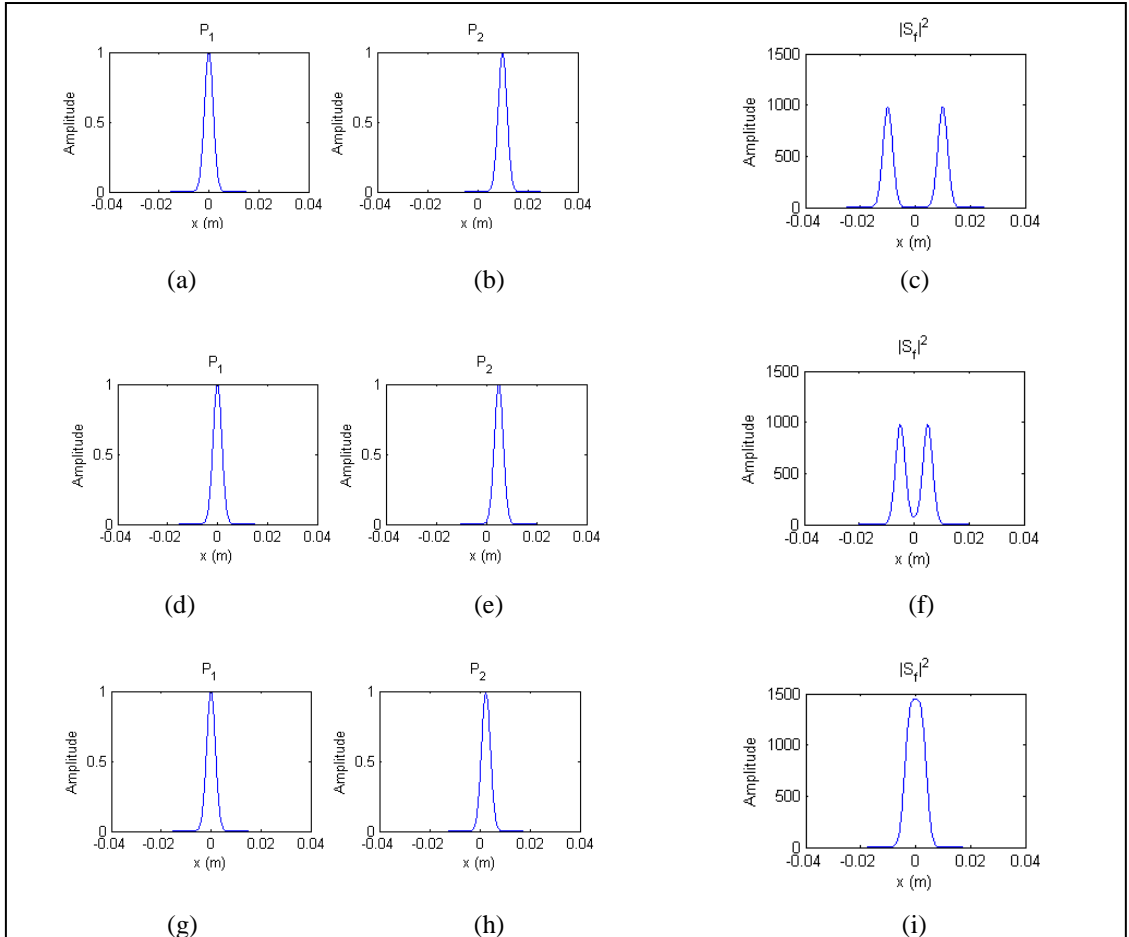

Fig 3: Illustration of the resolving power of the HOC architecture for one dimensional identical images

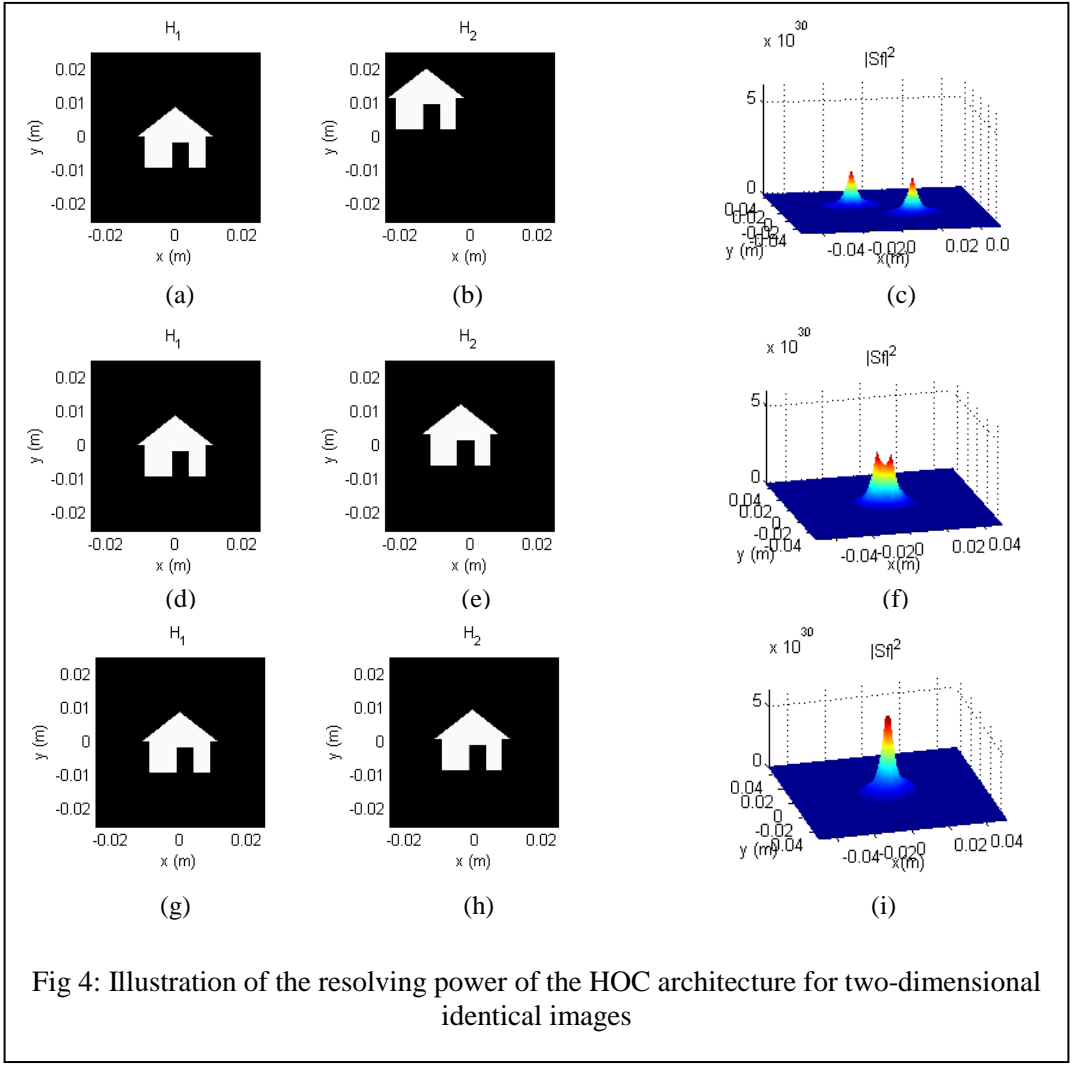

Fig 4: Illustration of the resolving power of the HOC architecture for two-dimensional identical images

Next, we consider some examples in two dimensions. Figures 4(a) and 4(b) show two identical images, shifted from each other by $|\vec{\rho_0}| \approx 0.02$ m. For this image $|\vec{\rho_m}| \approx 0.01$ m, so that $\eta > 1$ is satisfied, corresponding to case I. As can be seen in fig 4(c), the signal now has two distinct sharp peaks.

Next, we consider again the same images, but with a smaller shift: $|\vec{\rho_0}| \approx 0.004$ m, shown in figure 4(d) and 4(e). In this case $\eta \approx 0.4$, corresponding to case II, so that there will be overlaps between $T_3$ and $T_4$. However, the two peaks can still be discerned in the final signal, shown in fig 4(f). Finally we consider a case where, $|\vec{\rho_0}| = 0.001$ m, as shown in fig. 4(g) and 4(h). In this case, $\eta \approx 0.09$ and we can see that the peaks are no longer distinguishable, and have merged into each other, as shown in fig 4(i).

If the cross-correlation peaks are clearly resolved, then we can infer the distance between the two matched images, given by half the separation between the peaks. However, this information cannot be retrieved when the peaks are not resolved. On the other, the ability of the HOC architecture to determine whether a match is found is not adversely affected by the potential overlap between $T_3$ and $T_4$. Additional implications of this potential overlap between $T_3$ and $T_4$ are addressed in other sections of this paper. Note that in rest of the paper, we will assume that the convolution terms ($T_1$ and $T_2$) have been eliminated by the phase stabilization and scanning circuit.

## *B. Incorporating Polar Mellin Transform in the HOC architecture:*

If we start with regular reference and query images for the HOC architecture described above, it can detect images in shift invariant manner only. The architecture can be extended to achieve scale and rotation invariance, along with shift invariance, by transforming the reference and query images to log-polar domain. The flow diagram of performing this transformation, which is generally called the Polar Mellin Transform (PMT) is described in fig 5(a) and the detailed architecture is described in fig 5(b). We start with an image U(x',y'), (query or reference image captured by camera and converted to optical domain using an SLM as shown in fig 5(b)). The co-ordinates have dimensions of distance, e.g. meter. The next step is to find the FT of the image, $\tilde{U}(k_x', k_y')$ where the co-ordinates have dimensions of per meter. For notational convenience, we redefine $k_x' \rightarrow x$ and $k_y' \rightarrow y$, and denote as V(x,y) to be the same as $\tilde{U}(k_x', k_y')$. In practice, the original image, U(x',y') would be represented in the optical domain by using an SLM linked to a camera or a computer data base. A lens would be used to find the

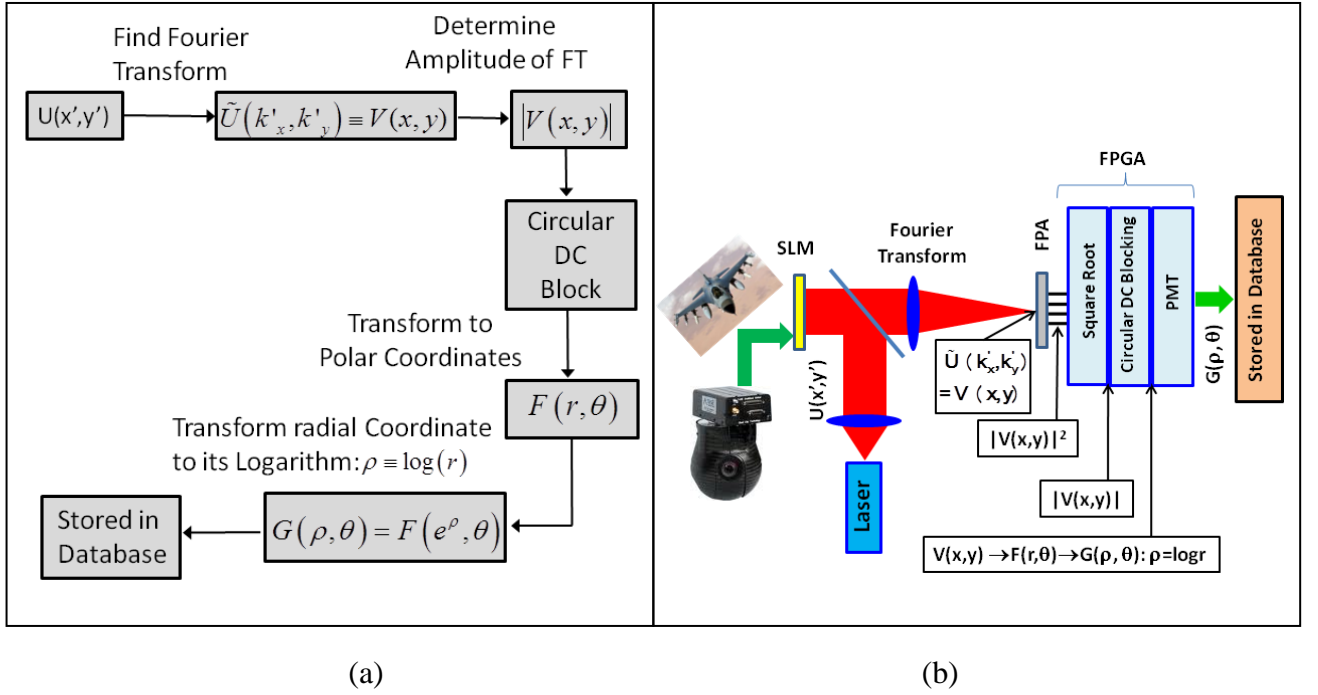

(a) (b)
Fig 5: (a) Flow diagram for transforming of query/ reference image to the log-polar domain.
(b) Schematic illustration of the architecture for implementing polar mellin transform.

fourier transform: $\tilde{U}(k'_x, k'_y) \equiv V(x, y)$. An FPA detects the intensity of the Fourier transformed image i.e. $|V(x,y)|^2$. The FPA is interfaced with the FPGA which determines the value of $|V(x,y)|$, thus eliminating the phase information. The magnitude of the FT of an object or function is invariant to a shift in the function $|F\{f(x,y)\}|=|F\{f(x-x_0)\}|$ but not to a scale change in the input. A circular hole of small radius (e.g. radius of 7 unit) on $|V(x,y)|$ is created using the FPGA. The necessity of creating this hole is discussed in detail in section 4, where we also point out that in general this hole does not affect significantly the performance of the correlator. To achieve scale and rotation invariance, the amplitude of $|V(x,y)|$ is transformed to the polar co-ordinate function, $F(r,\theta)$, using the FPGA. Then it is converted to the log-polar co-ordinate function $G(\rho,\theta)$ using the same FPGA. The reference images are Polar Mellin Transformed and stored in the database. The database can be a computer or a holographic memory disk if fast retrieval of the reference image is required. The captured query image also goes through the same procedure of PMT and supplied to the correlator's input port. In the next section, we describe the PMT process in detail, including illustrative examples.

# 3. Preprocessing the image using Polar Mellin Transform and examples of correlations using idealized images

A simple example of the PMT process is illustrated in figure 6. Here, we assume an artificial case where the amplitude of the FT [i.e., V(x,y)] of an image is choosen to be an uniform square, with a flower shape hole in it, as shown in fig. 6(a). The corresponding polar function, F(r,θ) is shown in fig. 6(b) and the corresponding polar- logarithmic function, G(ρ,θ) is shown in fig 6(c). Note that for F(r,θ), the coordinates *r* and *θ* are rectilinear (as opposed to curvilinear). For a given combination of coordinates in polar space, say $\{r = r_1; \theta = \theta_1\}$, we determine the corresponding values of *x* and *y* by using the relations $x_1 = r_1 \cos\theta_1$ and $y_1 = r_1 \sin\theta_1$. The value of the function F is then given by $F(r_1, \theta_1) = |V(x_1, y_1)|$. To plot the function F, we put the corresponding value of F at the co-ordinate ($x_1,y_1$) to a point that is a distance $r_1$ away from the origin along the horizontal axis (which has the dimension of length, mm) and a distance $\theta_1$ away from the origin along the vertical axis (which is in the dimension of radian, and span from 0 to $2\pi$).

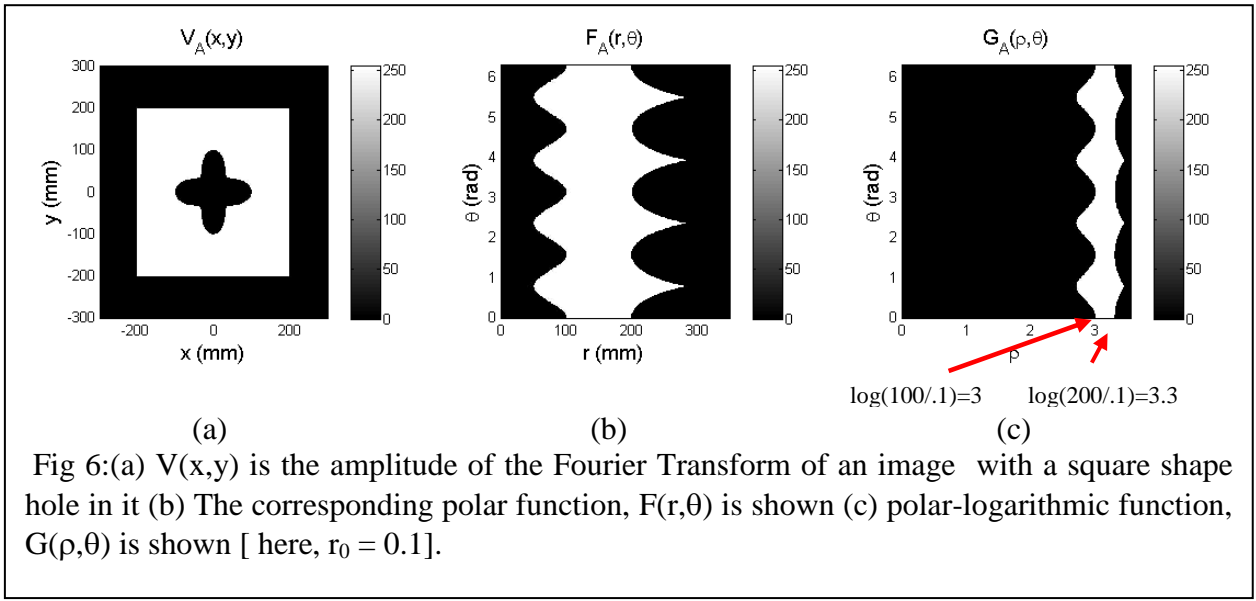

(a)      (b)      (c)

Fig 6:(a) V(x,y) is the amplitude of the Fourier Transform of an image with a square shape hole in it (b) The corresponding polar function, F(r,θ) is shown (c) polar-logarithmic function, G(ρ,θ) is shown [ here, $r_0 = 0.1$].

.

To generate the log-polar co-ordinate function $G(\rho,\theta)$, we proceed as follows, For a given combination of coordinates in this space, say $\{\rho = \rho_1; \theta = \theta_1\}$, we determine the corresponding values of r and θ (i.e. $r_1$ and $\theta_1$) by using the relations $\rho_1 = \log(r_1/r_0)$ and $\theta_1 = \theta_1$. Here the choice of the scaling distance, $r_0$, is arbitrary. Note that the value of $\log(r/r_0)$ approaches -∞ as r approaches zero, for any finite value of $r_0$. Obviously, this is an impractical situation. To circumvent this problem, we choose to ignore the information contained in a small circle of radius $r_0$ (in the V(x,y) plane), centered around r = 0, thus restricting the lower range of $\rho = \log(r/r_0)$ to 0, corresponding to r = $r_0$. The magnitude of $r_0$ should be chosen judiciously so as not to exclude any critical feature that may be present within the exclusion zone of $0 \leq r \leq r_0$. Of course, for the particular case shown in fig. 6, there is already a dark part in the center of $V_A(x,y)$. Thus, there is no loss of information if the circle of radius $r_0$ is fully contained in the small dark part. Figures 7(b) and 7(c) show the amplitude and the phase, respectively, of the image whose magnitude of FT is $V_A(x,y)$ as shown in fig 7(a). From fig 7(c) it is clear that such an image whose magnitude of the FT is $V_A(x,y)$ is unrealistic because the phase of the actual image is spanning between -π to π. While in this section, we restrict ourselves to unrealistic images where FT's have holes at the center, in section 4 we will consider realistic images for which it would be essential to exclude a small circle in FT plane.

Before proceeding further, we note that the necessity of excluding a small circle in the FT plane was not addressed in the previous investigations [9,10] pertaining to the use of PMT's for rotation and scale invariant correlations. This is due to in part to the fact that these papers [9,10] considered artificial images where FT's already contained holes in the center. Any realistic image, on the other hand, is bound to have a non-zero value at the center of the FT, corresponding to the average value of the image amplitude.

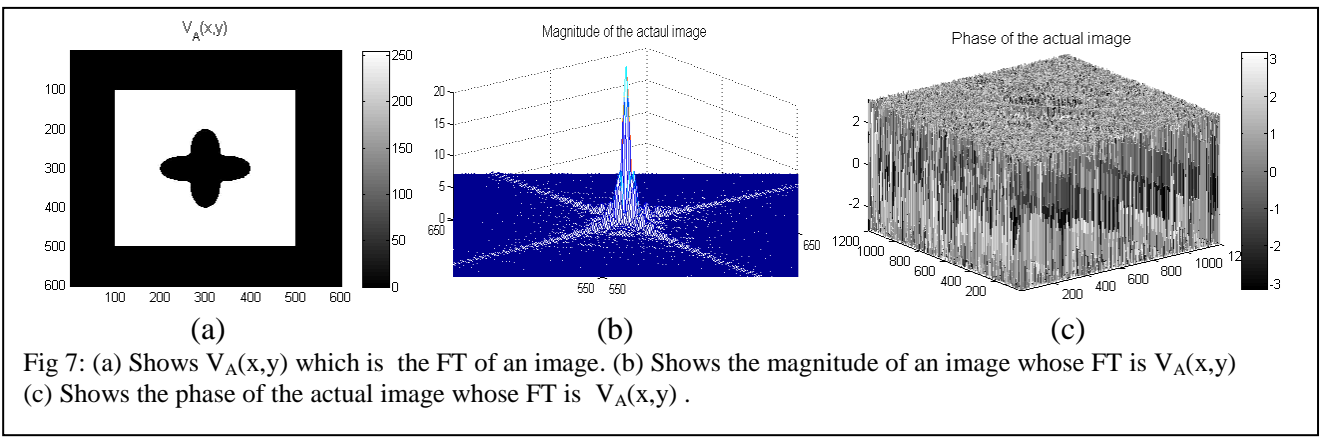

(a)      (b)      (c)

Fig 7: (a) Shows $V_A(x,y)$ which is the FT of an image. (b) Shows the magnitude of an image whose FT is $V_A(x,y)$ (c) Shows the phase of the actual image whose FT is $V_A(x,y)$.

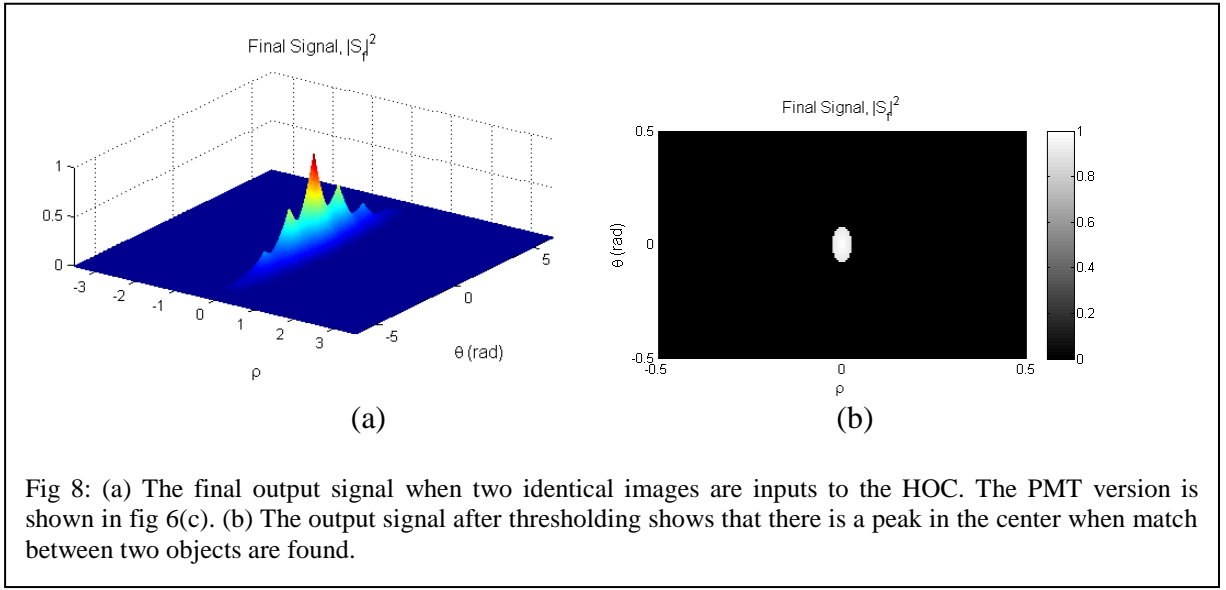

Fig 8: (a) The final output signal when two identical images are inputs to the HOC. The PMT version is shown in fig 6(c). (b) The output signal after thresholding shows that there is a peak in the center when match between two objects are found.

Consider now a situation where two identical PMT images, each corresponding to fig 6(c), are applied as inputs to the HOC. The resulting final output signals, $|S_f|^2$ of the HOC is illustrated in fig 8(a). Because of the perfect match the output has a sharp peak at the center, corresponding to the sum of the terms $T_3$ and $T_4$ (which are identical in this case) of eqn. 10. The value ($\sim 1$) at the peak, of course, is arbitrary, depending on the magnitude of the images. In fig 8(b), we show the peak clearly by applying a threshold value of 0.9.

Next we consider the effect of scale change on this PMT conversion process. Fig. 9 shows such a case, where $V_A(x,y)$ represents the amplitude of the FT of the same image as considered in fig.6, and $V_B(x,y)$ represents the amplitude of the FT of the same image, but scaled *up* by a linear factor of $\sigma = 2$ (i.e., x4 larger in area). Thus $V_B(x,y)$ is scaled *down* by a linear factor of 2 (x4 smaller in area) compared to $V_A(x,y)$. Note first that the corresponding polar distributions (as shown in fig 9(b) and 9(e)), $F_B(r,\theta)$, and $F_A(r,\theta)$ are the same in the $\theta$-direction, but differ by the linear scaling factor (2 in this case) in the r-direction. Note next that the corresponding polar-logarithmic distributions, $G_A(\rho,\theta)$ and $G_B(\rho,\theta)$ as shown in fig 9(c) and 9(f) respectively, are identical in shape, except for a shift in the $\rho$-direction, equaling the logarithm of the scale factor: $\log(2) \cong 0.3$. This illustrates the essence of how scale invariance is achieved via these transformations. The Fourier Transforms of $G_A(\rho,\theta)$ and $G_B(\rho,\theta)$ would be identical in amplitudes, thus leading to a strong cross-correlation when applied to the HOC , which eliminates the effect of relative shift.

Fig 9(g) and 9(h) illustrate the output signals of the HOC, if these PMT images [($G_A$ and $G_B$) or ($G_A$ and $G_C$)] are applied as inputs. The output signal of the correlator, $|S_f|^2$ contains the two cross-correlation terms, $T_3$ and $T_4$. From eqn 10, $T_3$ represents the cross-correlation of the PMT images, $G_A(\rho,\theta) \odot G_B(\rho,\theta)$ and $T_4$ represents the anti cross-correlation of the same PMT images, but in reverse order: $G_B(\rho,\theta) \odot G_A(\rho,\theta)$. Thus $T_3$ and $T_4$ signals are shifted by equal amount but in opposite direction according to the shift in position between $G_A(\rho,\theta)$ and $G_B(\rho,\theta)$. As shown in fig 9(g), the output now contains two peaks: a) the cross-correlation, $G_A(\rho,\theta) \odot G_B(\rho,\theta)$ at $\rho'= \log(2) = 0.3$ and $\theta' = 0$ and b) the anti cross-correlation $G_B(\rho,\theta) \odot G_A(\rho,\theta)$ at $\rho'= -\log(2) = -0.3$ and $\theta' = 0$, where $(\rho',\theta')$ are the co-ordinates in the correlation plane. Here, as expected, the magnitude of each peak is ~0.27, which is approximately one fourth of the peak value shown in fig. 8. In figure 9(h), we apply a threshold of 0.25 to illustrate the peaks clearly.

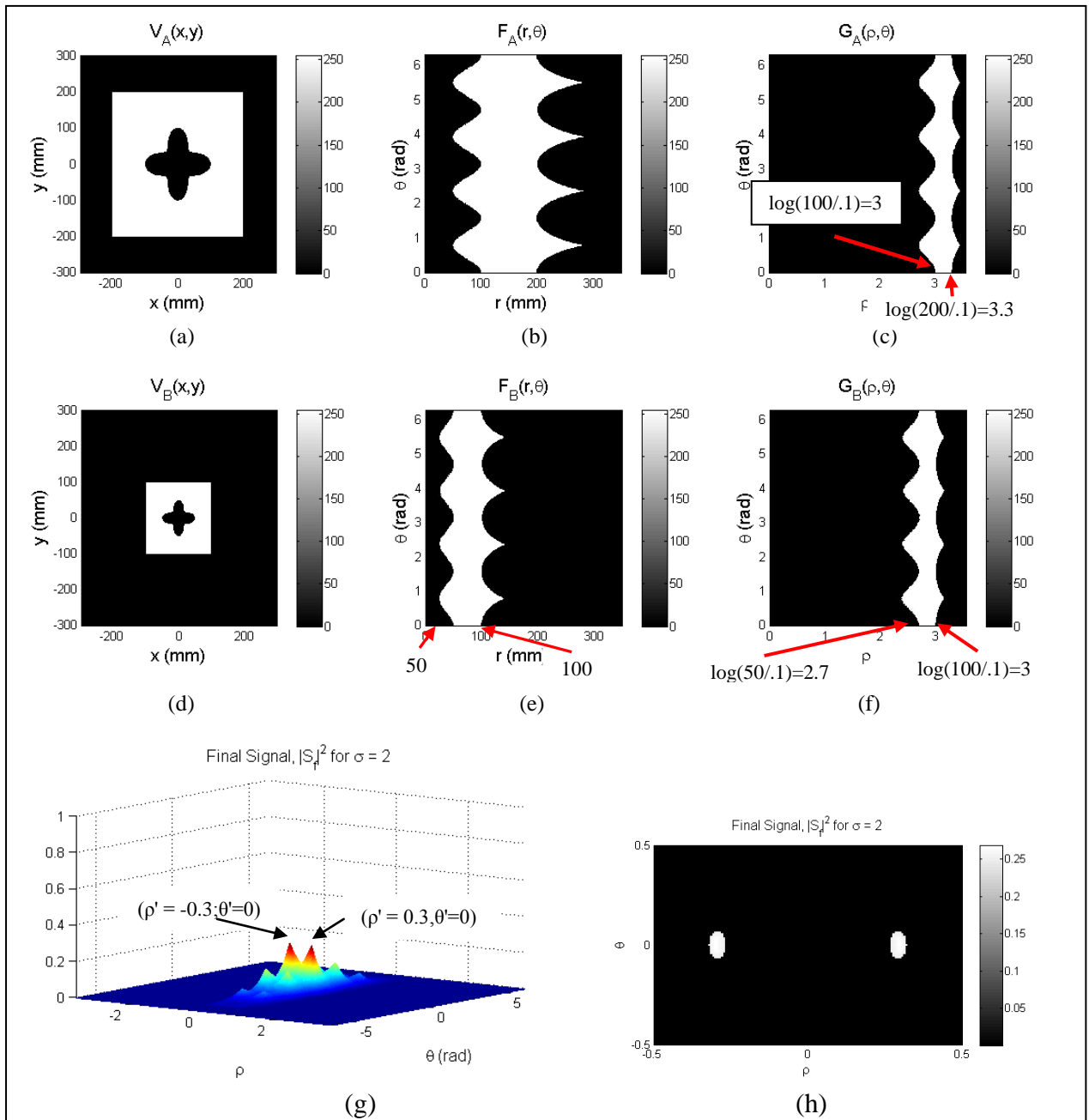

Fig 9: (a) We consider an artificial case where, $V_A(x,y)$ is the FT of an image (b) the corresponding polar distribution $F_A(r,\theta)$ (c) the corresponding log-polar distribution $G_A(\rho,\theta)$ (d) $V_B(x,y)$ is smaller in area than $V_A(x,y)$ by a factor of 4. (e) the corresponding polar distributions, $F_B(r,\theta)$ and $F_A(r,\theta)$, are the same in the $\theta$-direction, but differ by the linear scaling factor (2 in this case) in the r-direction (f) the corresponding polar-logarithmic distribution, $G_B(\rho,\theta)$. $G_A(\rho,\theta)$ and $G_B(\rho,\theta)$ are identical in shapes, except for a shift in the $\rho$-direction [equaling the logarithm of the scale factor: $\log(2) \cong 0.3$]. (a) $G_A(\rho,\theta)$ and $G_B(\rho,\theta)$ is applied to the correlator ; The final signal of $G_A(\rho,\theta)$ and $G_B(\rho,\theta)$ has two peaks shifted in the $\rho$-direction by an amount of $\log(2) = 0.3$. (b) The final signal after thresholding for $\sigma = 2$.

It is important to consider the limit imposed on this process by the fact that the final signal contains both $T_3$ and $T_4$ terms. As we discussed in detail in section 2, there are essentially two distinct scenarios, characterized by the parameter $\eta$. For the scale invariant recognition, the value of this parameter is proportional to the scaling factor. The case shown in fig 9 corresponds to $\eta > 1$, producing two peaks that are clearly resolved. We next consider a case where the scaling factor is small, corresponding to $\eta < 1$. In this case, $V_C(x,y)$ is scaled *down* by a small factor of $\sigma = 1.2$ compared to $V_A(x,y)$ as shown in fig 10(a). The corresponding polar distribution and log-polar distributions $F_C(r,\theta)$ and $G_C(\rho,\theta)$, are shown in fig 10(b) and 10(c), respectively. For this case, the polar-logarithmic distributions, $G_A(\rho,\theta)$ and $G_C(\rho,\theta)$ are shifted from one another by an amount $\log(1.2) \approx 0.08$ which corresponds to $\eta < 1$.

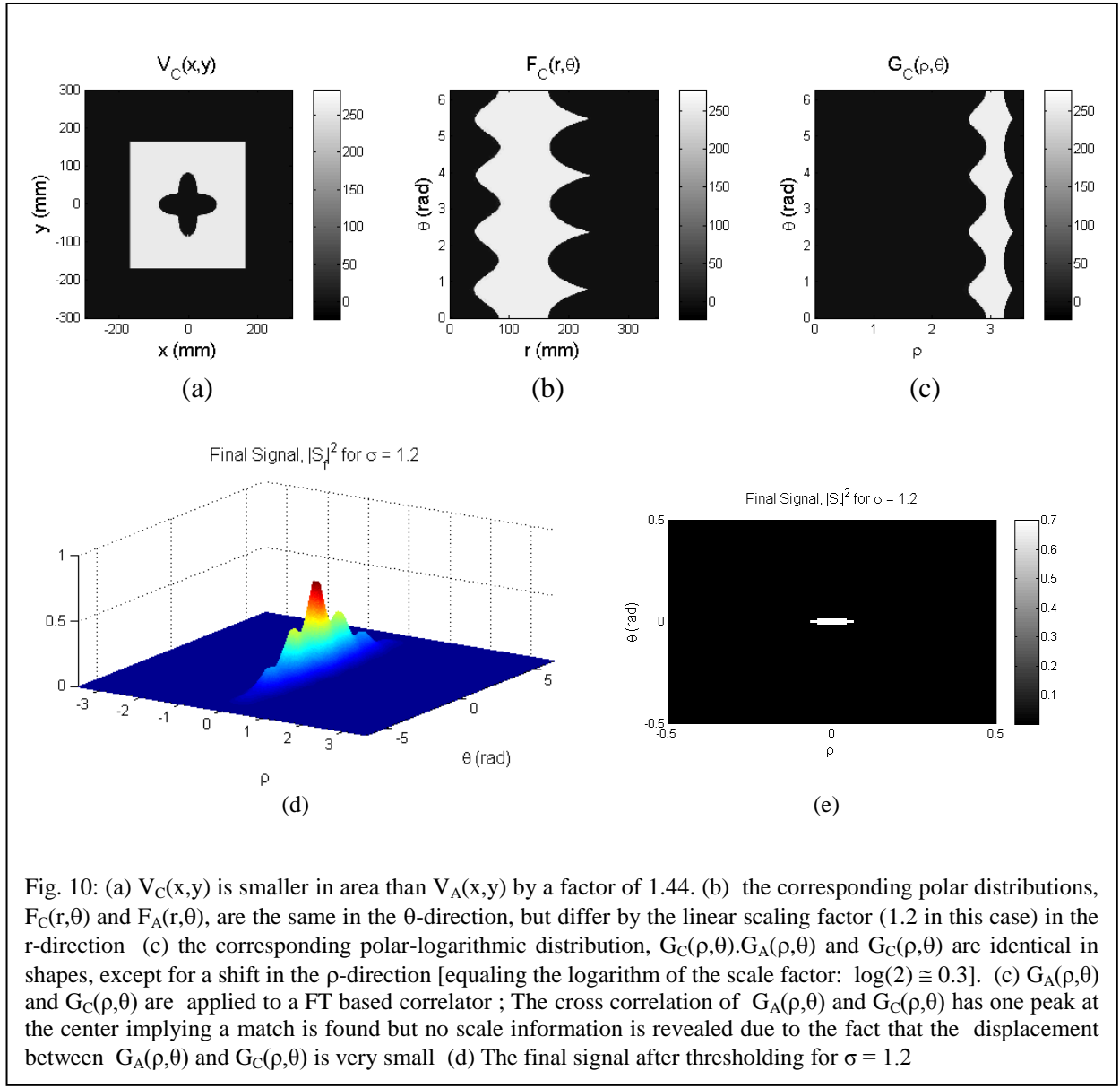

Fig. 10: (a) $V_C(x,y)$ is smaller in area than $V_A(x,y)$ by a factor of 1.44. (b) the corresponding polar distributions, $F_C(r,\theta)$ and $F_A(r,\theta)$, are the same in the θ-direction, but differ by the linear scaling factor (1.2 in this case) in the r-direction (c) the corresponding polar-logarithmic distribution, $G_C(\rho,\theta).G_A(\rho,\theta)$ and $G_C(\rho,\theta)$ are identical in shapes, except for a shift in the ρ-direction [equaling the logarithm of the scale factor: $\log(2) \cong 0.3$]. (c) $G_A(\rho,\theta)$ and $G_C(\rho,\theta)$ are applied to a FT based correlator ; The cross correlation of $G_A(\rho,\theta)$ and $G_C(\rho,\theta)$ has one peak at the center implying a match is found but no scale information is revealed due to the fact that the displacement between $G_A(\rho,\theta)$ and $G_C(\rho,\theta)$ is very small (d) The final signal after thresholding for σ = 1.2

Fig 10(d) shows the final signal while $G_A(\rho,\theta)$ and $G_C(\rho,\theta)$ are applied as input the HOC architecture. In this case, the corresponding shift between $G_A(\rho,\theta)$ and $G_C(\rho,\theta)$ is $\log(1.2) \approx 0.08$. Hence, the two peaks have merged into a single peak of magnitude ∼ 0.7. Note that in this case, while we are still able to determine the fact that the images are matched, the information about the relative scale between the images is lost. Fig 10(e) shows the final output signal after thresholding.

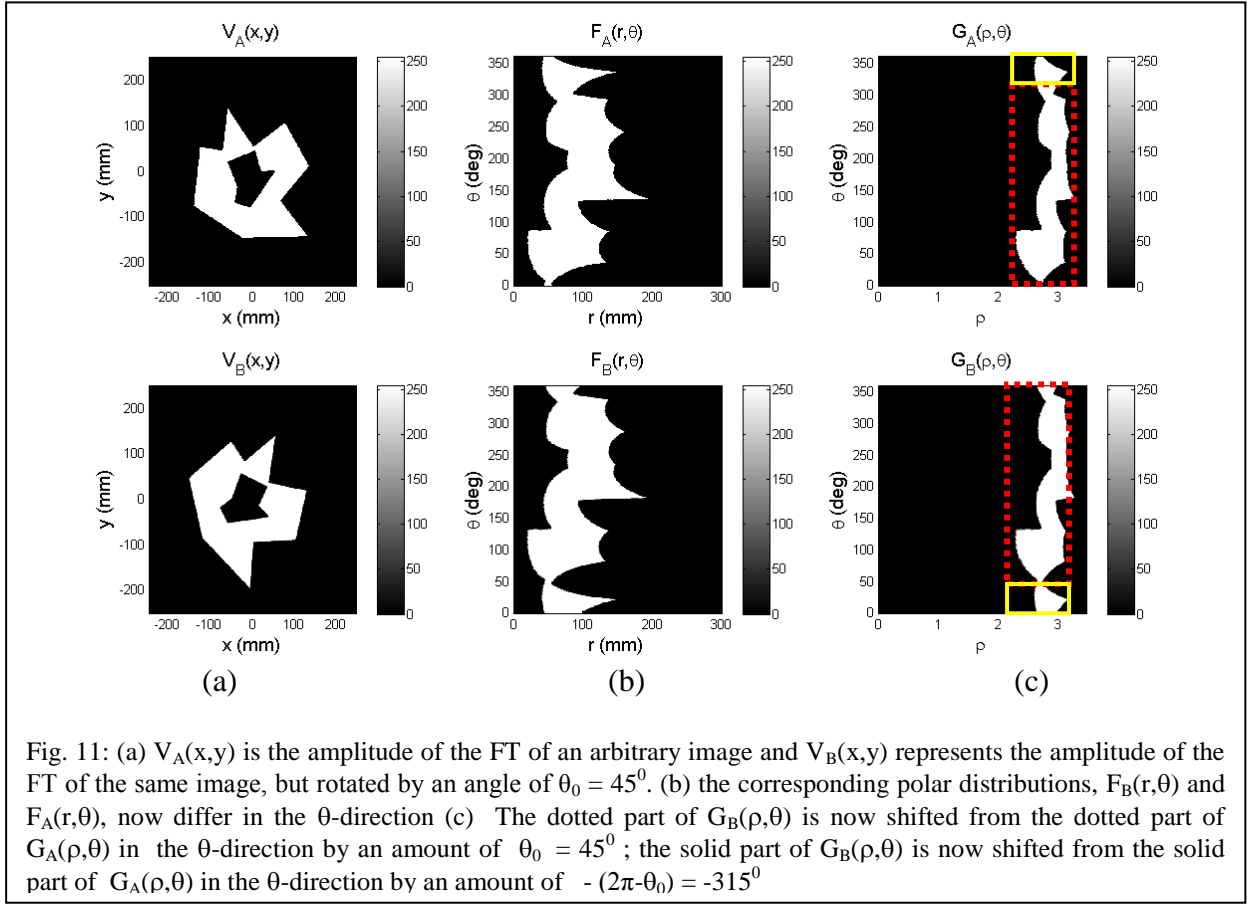

Fig. 11: (a) $V_A(x,y)$ is the amplitude of the FT of an arbitrary image and $V_B(x,y)$ represents the amplitude of the FT of the same image, but rotated by an angle of $\theta_0 = 45^0$. (b) the corresponding polar distributions, $F_B(r,\theta)$ and $F_A(r,\theta)$, now differ in the $\theta$-direction (c) The dotted part of $G_B(\rho,\theta)$ is now shifted from the dotted part of $G_A(\rho,\theta)$ in the $\theta$-direction by an amount of $\theta_0 = 45^0$; the solid part of $G_B(\rho,\theta)$ is now shifted from the solid part of $G_A(\rho,\theta)$ in the $\theta$-direction by an amount of $-(2\pi-\theta_0) = -315^0$

Now, we consider the effect of rotation on the PMT images. Fig 11 shows such a case, where $V_A(x,y)$ is the amplitude of the FT of another arbitrary image and $V_B(x,y)$ represents the amplitude of the FT of the same image, but rotated by an angle of $\theta_0 = 45^0$. Thus, $V_B(x,y)$ is rotated by an angle of $\theta_0 = 45^0$ compared to $V_A(x,y)$. Here we have made use of the well-known fact that the process of FT preserves the angular information. Note first that the corresponding polar distributions (as shown in fig 11(b)), $F_B(r,\theta)$, and $F_A(r,\theta)$ now differ in the $\theta$-direction only and the pattern is shifted by $\theta_0 = 45^0$. Also note that the log-polar distributions are now still identical in shapes, except that $G_B(\rho,\theta)$ is now shifted from $G_A(\rho,\theta)$ in the $\theta$-direction. It is particularly important to consider the shift in the $\theta$-direction very carefully, since this coordinate for $\rho = 0$) of the final signal. The output signal contains several peaks, corresponding to the cross-correlation and anti corss correlation terms, $T_3$ and $T_4$, of eqn. 9, which is shown in fig 12(a) and (b). Consider first the $T_3$ term, which corresponds to $G_A(\rho,\theta) \odot G_C(\rho,\theta)$. Since there is no match between $G_{A1}(\rho,\theta)$ and $G_{B2}(\rho,\theta)$, and between $G_{A2}(\rho,\theta)$ and $G_{B1}(\rho,\theta)$, we get only two peaks: peak 3a corresponding to $G_{A1}(\rho,\theta) \odot G_{B1}(\rho,\theta)$ at $\rho'=0$ and $\theta'=\theta_0$ (=$45^0$), and peak 3b corresponding to $G_{A2}(\rho,\theta) \odot G_{B2}(\rho,\theta)$ at $\rho' =0$ and $\theta'= - (2\pi - \theta_0) = -315^0$. The signal corresponding to $T_3$, for $\rho = 0$, is plotted as a function of $\theta$ in figure 12(c). As can be seen, peak 3a is prominent, while peak 3b is barely visible. This is due to the fact that $\theta_0$ is small compared to $(2\pi - \theta_0)$ so that the energy contained in $G_{A2}$ ($G_{B2}$) is smaller than that contained in $G_{A1}$ ($G_{B1}$). Similarly, from the $T_4$ term, we get two other peaks: peak 4a corresponding to $G_{B1}(\rho,\theta) \odot G_{A1}(\rho,\theta)$ at $\rho' = 0$ and $\theta' = -\theta_0$; and peak 4b corresponding to $G_{B2}(\rho,\theta) \odot G_{A2}(\rho,\theta)$ at $\rho'= 0$ and $\theta' = (2\pi - \theta_0)$. These two peaks are shown in fig 12(d), as a function of $\theta$, for $\rho' = 0$. Again, we see that peak 4b is much smaller than peak 4a. The final output signal $|S_f|^2$ is shown in fig 12 (e) where the locations of the peaks can is clearly visible after thresholding (shown in fig 12(f)). In fig 13, we show how the relative amplitudes of peak 3a and 3b vary as a function of the rotation angle, $\theta_0$. Similar behavior occurs (not shown) for peak 4a and peak 4b as well. The actual ratios of these peaks would, of course, depend on the details of the angular properties of the image. In fig. 13(b), we see peaks 3a and 4a clearly, while peaks 3b and 4b barely visible. However, the detection of just two peaks is enough to discern the relative angle of rotation.

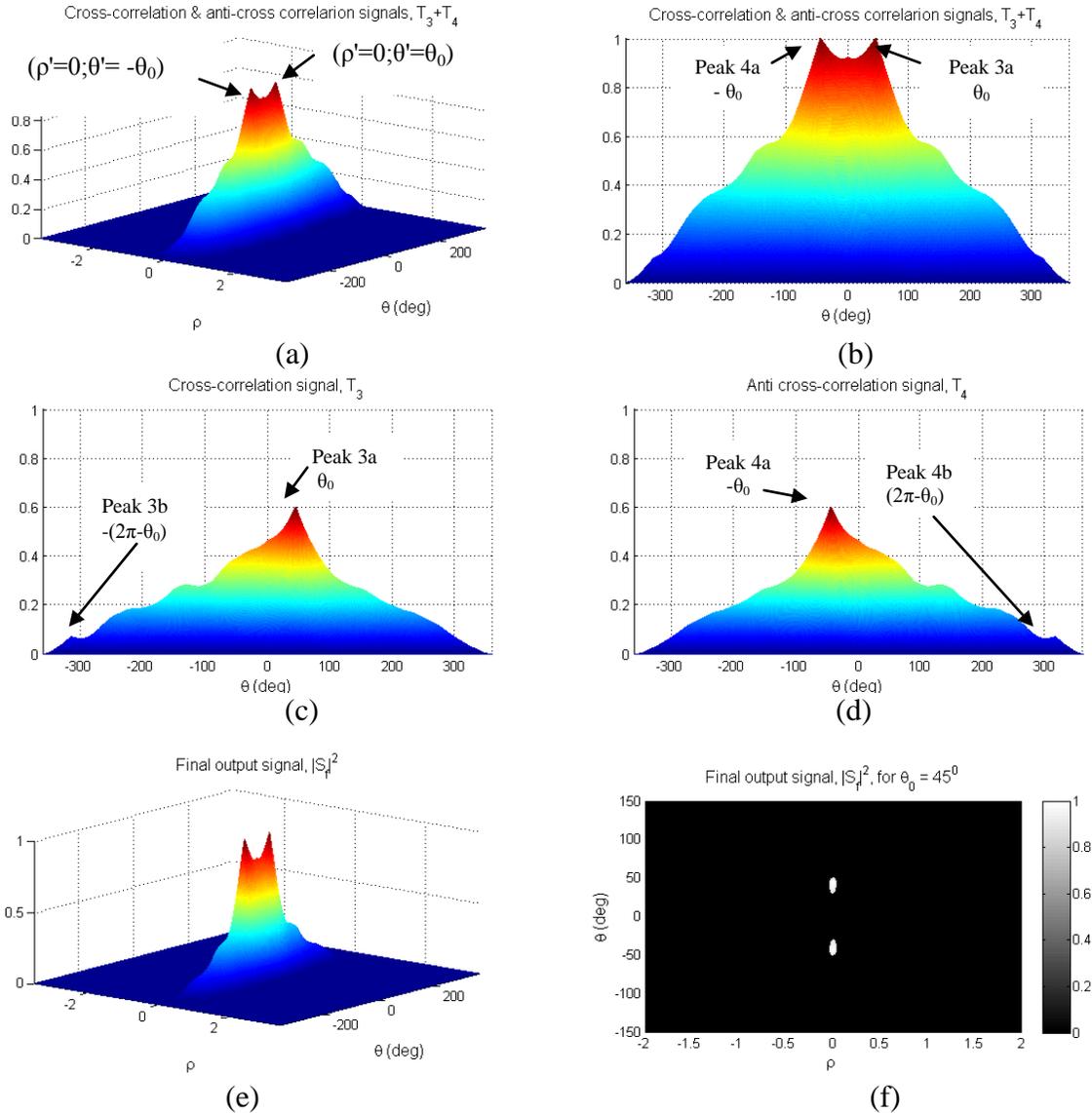

Fig. 12: (a) The final output when $G_A(\rho,\theta)$ and $G_C(\rho,\theta)$ are applied to the HOC; (b) The cross-sectional view of the final signal showing two peaks shifted in the $\theta$-direction by an amount of $\theta_0$ and $-\theta_0$, which correspond to the cross-correlation terms $G_{A1}(\rho,\theta) \odot G_{C1}(\rho,\theta)$ and $G_{C1}(\rho,\theta) \odot G_{A1}(\rho,\theta)$, respectively. (c) The cross-correlation signal $T_3$ which shows the two peaks corresponding to the cross-correlation $G_{A1}(\rho,\theta) \odot G_{C1}(\rho,\theta)$ and $G_{A2}(\rho,\theta) \odot G_{C2}(\rho,\theta)$. (d) The anti cross-correlation signal $T_3$ which shows the two peaks corresponding to the cross-correlations $G_{C1}(\rho,\theta) \odot G_{A1}(\rho,\theta)$ and $G_{C2}(\rho,\theta) \odot G_{A2}(\rho,\theta)$. (e) The final output signal $|S_f|^2$. (f) The final output signal after thresholding.

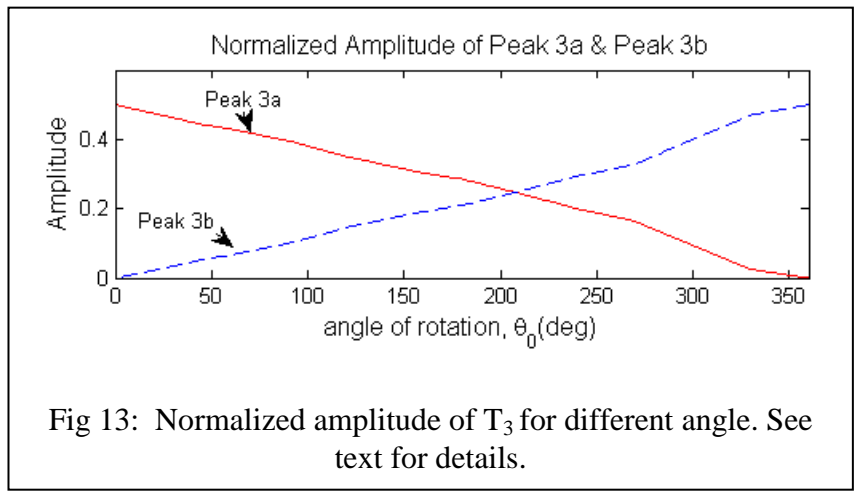

Fig 13: Normalized amplitude of $T_3$ for different angle. See text for details.

Next, we consider the effect of rotation and scale change simultaneously. Fig 14 shows such a case, where $V_A(x,y)$ is the amplitude of the FT of an arbitrary image and $V_C(x,y)$ represents the amplitude of the FT of the same image, but scaled *up* by a linear factor of 2 and rotated by an angle of $\theta_0 = 45^0$. Note first that the corresponding polar distributions (as shown in fig. 14(b)), $F_C(r,\theta)$, and $F_A(r,\theta)$ now differ in both the r-direction and the $\theta$-direction. Note next that the log-polar distributions are now still identical in shapes, except that $G_C(\rho,\theta)$ is now shifted from $G_A(\rho,\theta)$ in the $\rho$-direction by an amount $\rho_0 = \log(2) = 0.3$ and the $\theta$-direction by an amount $\theta_0 = 45^0$. Similar to the case of rotation change described above, the distribution along the $\theta$-direction in $G_A$ ($G_C$) can be broken into two parts: the part enclosed in the solid box in the upper (lower) part of fig 14(c) can be denoted as $G_{A1}(G_{C1})$ and the part enclosed in the dotted box is denoted as $G_{A2}$ ($G_{C2}$). In $G_C(\rho,\theta)$ the $G_{C1}$ part is shifted by an angle $\theta_0$, while the $G_{C2}$ part is shifted by $-(2\pi-\theta_0)$, compared to $G_{A1}$ and $G_{A2}$, respectively.

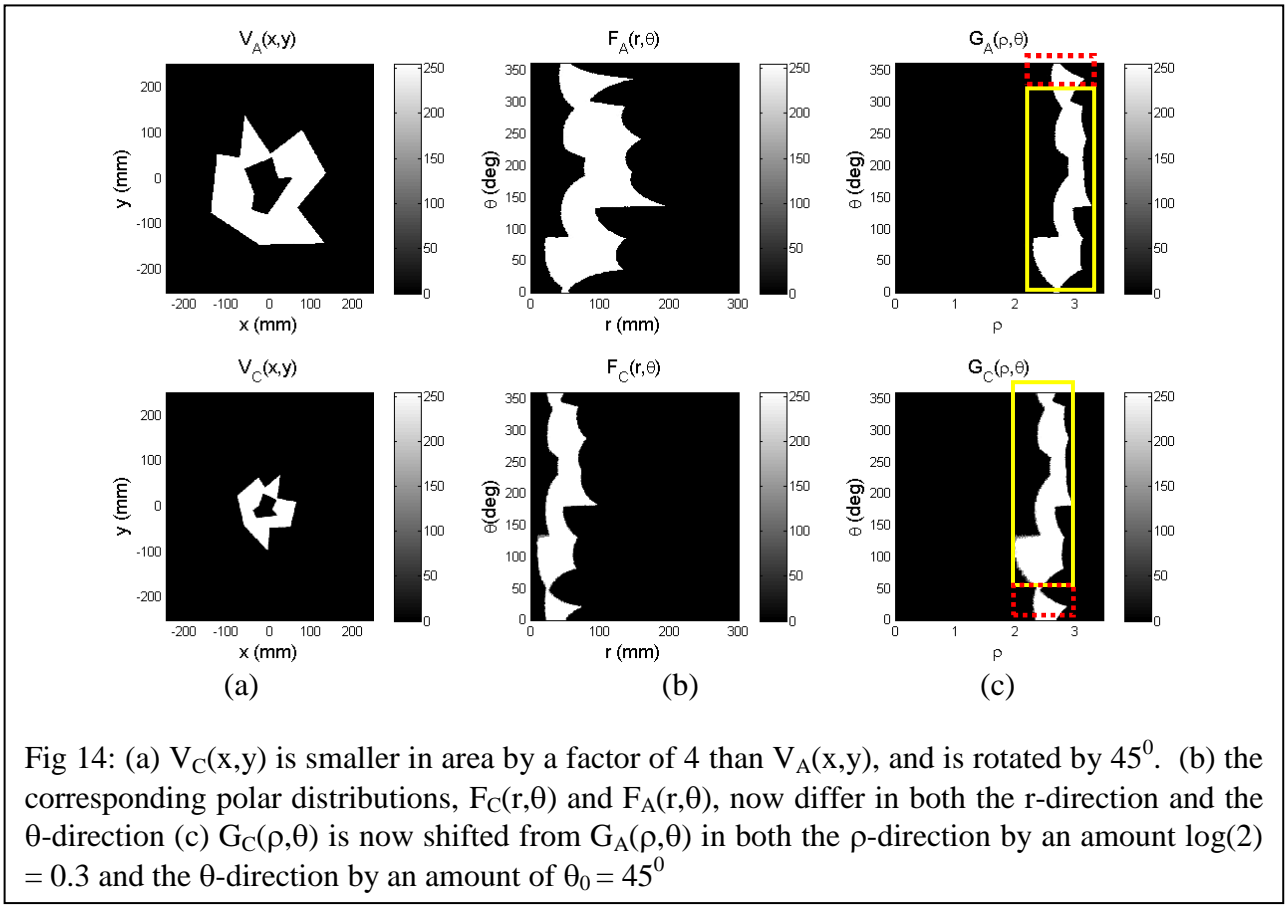

Fig 14: (a) $V_C(x,y)$ is smaller in area by a factor of 4 than $V_A(x,y)$, and is rotated by $45^0$. (b) the corresponding polar distributions, $F_C(r,\theta)$ and $F_A(r,\theta)$, now differ in both the r-direction and the $\theta$-direction (c) $G_C(\rho,\theta)$ is now shifted from $G_A(\rho,\theta)$ in both the $\rho$-direction by an amount $\log(2) = 0.3$ and the $\theta$-direction by an amount of $\theta_0 = 45^0$

Fig 15(a) shows the output signal, when $G_A(\rho,\theta)$ and $G_C(\rho,\theta)$ are applied to the HOC. The output contains several peaks, corresponding to the cross-correlation and anti cross correlation terms, $T_3$ and $T_4$, of eqn. 9. Consider first the $T_3$ term, which corresponds to $G_A(\rho,\theta) \odot G_C(\rho,\theta)$. Since there is no match between $G_{A1}(\rho,\theta)$ and $G_{C2}(\rho,\theta)$, and between $G_{A2}(\rho,\theta)$ and $G_{C1}(\rho,\theta)$, we get only two strong peaks: peak 3a' corresponding to $G_{A1}(\rho,\theta) \odot G_{C1}(\rho,\theta)$ at $\rho'= \log(2) = 0.3$ and $\theta' = \theta_0$ ($_= 45^0$); and peak 3b' corresponding to $G_{A2}(\rho,\theta) \odot G_{C2}(\rho,\theta)$ at $\rho' = 0.3$ and $\theta' = -(2\pi-\theta_0)$ (shown in fig. 15(c)). As can be seen, peak 3a' is prominent, while peak 3b' is barely visible. Similarly, from the $T_4$ term, we get two other peaks: peak 4a' corresponds to $G_{C1}(\rho,\theta) \odot G_{A1}(\rho,\theta)$ at $\rho' = -0.3$ and $\theta' = -\theta_0$; and peak 4b' corresponds to $G_{C2}(\rho,\theta) \odot G_{A2}(\rho,\theta)$ at $\rho'= -0.3$ and $\theta' = (2\pi-\theta_0)$ (shown in fig. 15(d)). Fig 15(e) shows the final output signal $|S_f|^2$ and fig 15(f) shows the output signal after applying a threshold of 0.9. From fig 15(f), it is obvious that the positions of the peaks of the Cross-correlation signals correspond to rotation and scale change between two images.

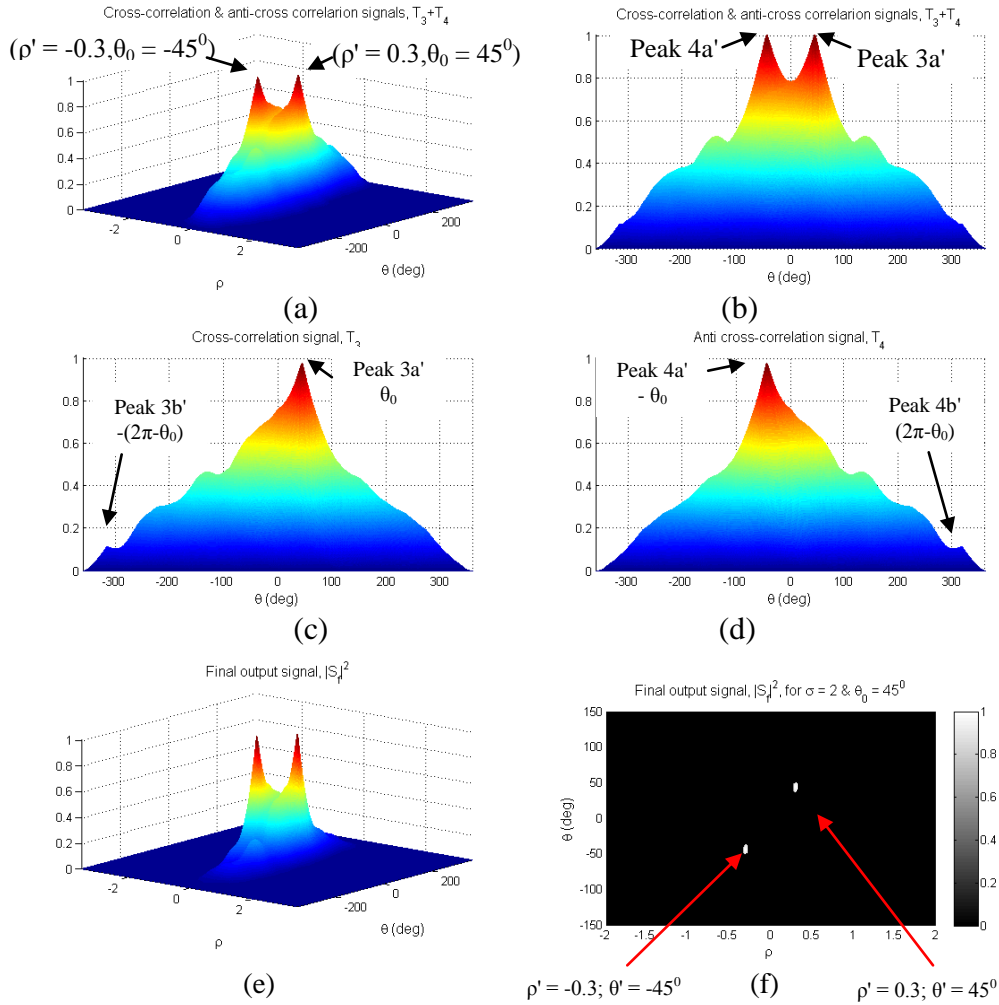

Fig. 15: (a) The final output when $G_A(\rho,\theta)$ and $G_C(\rho,\theta)$ are applied to the HOC, showing two peaks shifted in $\theta$-direction by an amount of $\theta_0$ and $-\theta_0$. These correspond to the cross-correlation terms $G_{A1}(\rho,\theta)\odot G_{C1}(\rho,\theta)$ and $G_{C1}(\rho,\theta)\odot G_{A1}(\rho,\theta)$, respectively. (b) The cross-sectional view shows the peaks 3a' and 4a' (c) The cross-correlation signal $T_3$ which shows the two peaks corresponding to the cross-correlation $G_{A1}(\rho,\theta)\odot G_{C1}(\rho,\theta)$ and $G_{A2}(\rho,\theta)\odot G_{C2}(\rho,\theta)$. (d) The anti cross-correlation signal $T_3$ which shows the two peaks corresponding to the cross-correlations $G_{C1}(\rho,\theta)\odot G_{A1}(\rho,\theta)$ and $G_{C2}(\rho,\theta)\odot G_{A2}(\rho,\theta)$ (e) The final output signal $|S_f|^2$. (f) The final output signal after thresholding.

## 4. Simulation Results of the Scale and Rotation invariant HOC

In section 3, we have shown that by incorporating Polar Mellin transform, the proposed HOC architecture can achieve scale and rotation invariance, in addition to the shift invariance feature. So far, we have considered artificial cases where the FT of the object or reference image has a distinct hole in the center. As discussed earlier this corresponds to an unrealistic situation where the image must have both positive and negative amplitudes. In this section, we consider real world scenarios where the image has non-negative values everywhere. Since such an image has a non-zero average value, its FT cannot have a hole in the center. By cutting a hole of a suitable radius in the center of the FT of the image, we produce an effective image, which is no longer positive definite, and is thus compatible with the PMT process.

In fig. 16(a), we consider two images, $U_1(x',y')$ and $U_2(x',y')$, where $U_2(x',y')$ is scaled down by a linear factor of 2 with respect to $U_1(x',y')$. The corresponding magnitudes of FTs, $V_1(x,y)$ and $V_2(x,y)$, are shown in fig 16(b), where $V_2(x,y)$ is scaled up by a factor of 2 compared to $V_1(x,y)$ in each dimension. Here, we take the magnitude of the FTs to get rid of any shift information. In fig 16(c), a hole of radius $r_0$ is created in the center of $V_1(x,y)$ and $V_2(x,y)$, which are denoted as $V_{1H}(x,y)$ and $V_{2H}(x,y)$, respectively. Fig 16(d) shows the corresponding polar distributions, $F_1(r,\theta)$ and $F_2(r,\theta)$. As shown in fig 16(e), the polar-logarithmic distributions are nearly identical in shape, except that $G_2(\rho,\theta)$ is now shifted from $G_1(\rho,\theta)$ in the $\rho$-direction by an amount of $\log(2) = 0.3$. The PMT processed images $G_1(\rho,\theta)$ and $G_2(\rho,\theta)$, are now inputs to the HOC architecture. Fig 17(a) shows the final output signals $|S_f|^2$ of the HOC where the peaks of the cross-correlation signals, $T_3$ and $T_4$ are shifted from the center in the $\rho$ direction by an amount equaling $\log(\sigma)$ and $-\log(\sigma)$, respectively. Fig 17(b) shows that thresholding gives a clear view of the location of these peaks, from which we can determine the relative scaling factor.

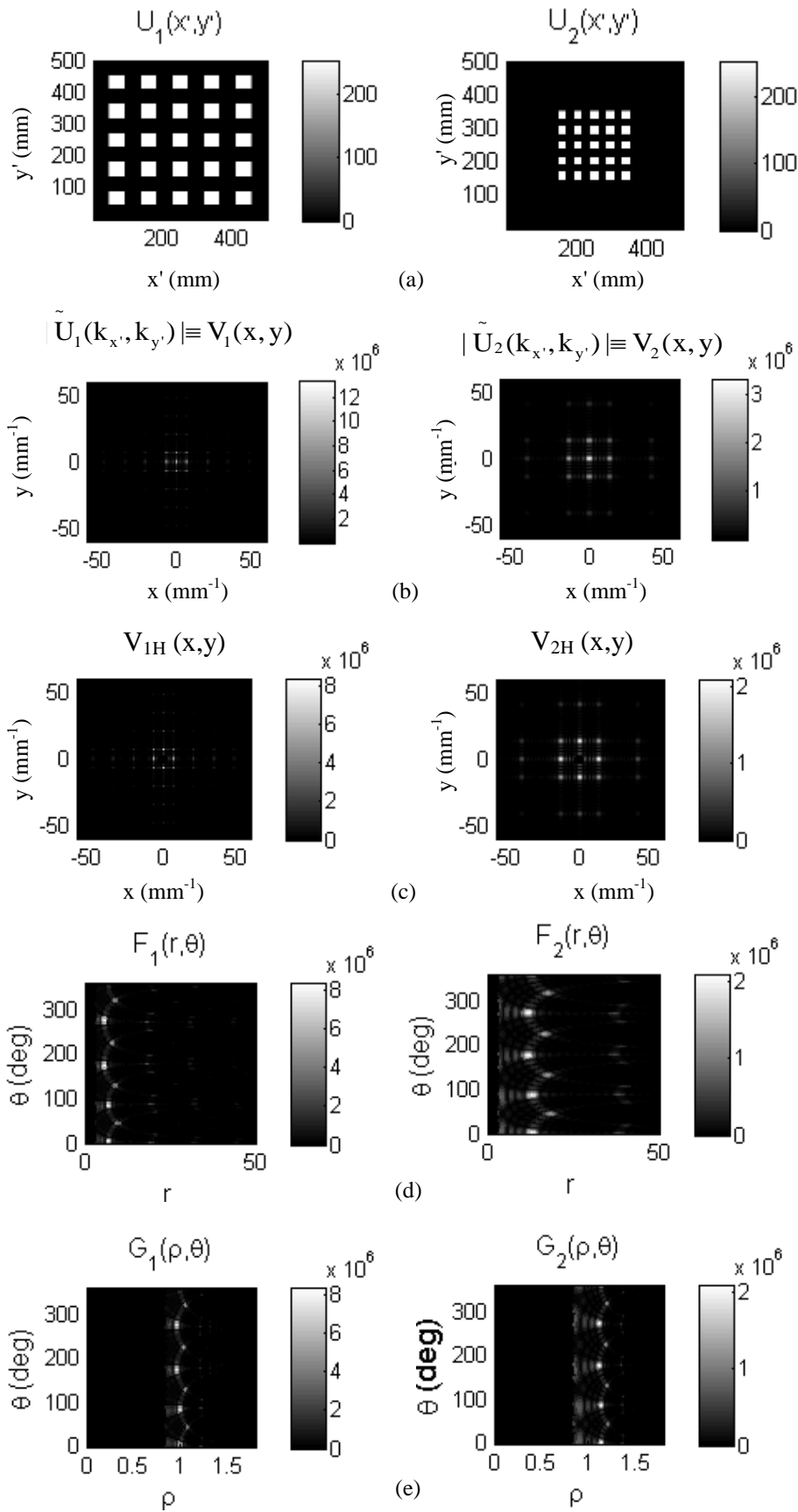

Fig. 16:(a) we consider two images $U_1(x',y')$ and $U_2(x',y')$, where $U_2(x',y')$ is smaller in area than $U_1(x',y')$ by a factor of 4; (b) The corresponding FTs are denoted as $V_1(x,y)$ and $V_2(x,y)$, which are similar in shape except that the FT of $V_2(x,y)$ has a larger area than that of $V_1(x,y)$ by a factor of 4; (c) A hole of radius $r_0$ is created in the center of $V_1(x,y)$ and $V_2(x,y)$ and the resulting function are denoted as $V_{1H}(x,y)$ and $V_{2H}(x,y)$, respectively; (d) The corresponding polar distributions, $F_1(r,\theta)$ and $F_2(r,\theta)$, are the same in the θ-direction, but differ by a linear scaling factor of 2 in the r-direction. (e) The polar-logarithmic distributions, $G_1(\rho,\theta)$ and $G_2(\rho,\theta)$, are identical in shapes, except for a shift in the ρ-direction [equaling the logarithm of the scale factor: $\log(2) \cong 0.3$]

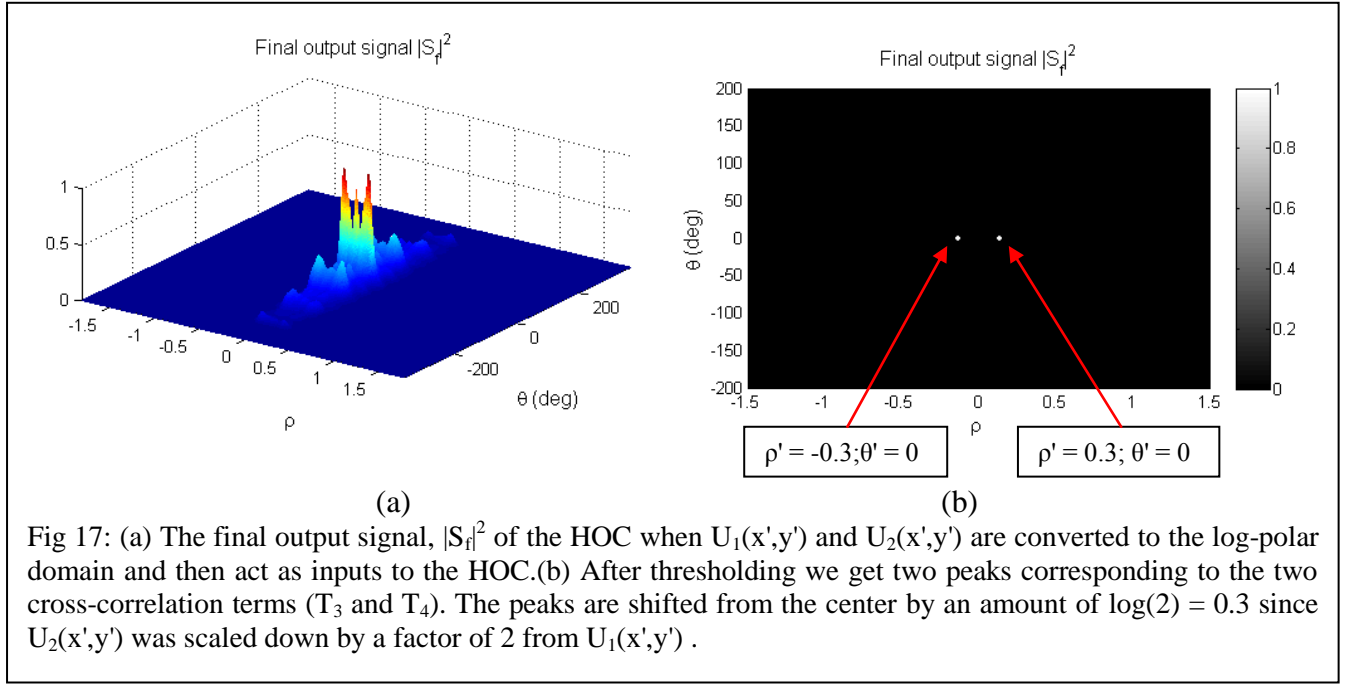

Fig 17: (a) The final output signal, $|S_f|^2$ of the HOC when $U_1(x',y')$ and $U_2(x',y')$ are converted to the log-polar domain and then act as inputs to the HOC.(b) After thresholding we get two peaks corresponding to the two cross-correlation terms ($T_3$ and $T_4$). The peaks are shifted from the center by an amount of $\log(2) = 0.3$ since $U_2(x',y')$ was scaled down by a factor of 2 from $U_1(x',y')$.

When a hole is cut in the FT of an image, the process is equivalent to the use of a modified image. We use one of the images considered above $U_1(x',y')$, above to illustrate what this modified image looks like. Before preceding it is instructive to document clearly the notations we have employed; as shown in table 1.

| Symbol | Meaning |
| --- | --- |
| $U_1(x',y')$ | Original image; {x',y'} are spatial coordinates, with units of meter (m) |
| $\tilde{U}_1(k_{x'}, k_{y'})$ | FT of the original image; {$k_{x'}$, $k_{y'}$} are wave number coordinates, with units of m$^{-1}$ |
| $\tilde{U}_1(x, y)$ | Same as $\tilde{U}_1(k_{x'}, k_{y'})$, except that we have defined $x \equiv k_{x'}$, $y \equiv k_{y'}$; Thus, {x,y} are wave number coordinates, with units of mm$^{-1}$. This redefinition is for convenience only. |
| $V_1(x,y) \equiv |\tilde{U}_1(x, y)|$ | This is the magnitude of the FT of original image. |
| $U_{1H}(x',y')$ | Image resulting from cutting a hole in the FT. |
| $\tilde{U}_{1H}(k_{x'}, k_{y'})$ | FT of $U_{1H}(x',y')$; Again the original image; {x',y'} are spatial coordinates. |

| $\tilde{U}_{1H}(x, y)$ | Same as $\tilde{U}_{1H}(k_{x'}, k_{y'})$ with the definition of $x \equiv k_{x'}$, $y \equiv k_{y'}$ |
|---|---|
| $V_{1H}(x,y) \equiv |\tilde{U}_{1H}(x,y)|$ | This is the magnitude of the FT of the modified image. |

Table 1: Summary of definitions of various transform

In the correlation process described in fig. 17, we made use of $V_{1H}(x,y)$ and $V_{2H}(x,y)$. The corresponding modified images are $U_{1H}(x',y')$ and $U_{2H}(x',y')$. As an example, we show below, in steps, how to determine $U_{1H}(x',y')$, and explain its shape.

Fig. 18(a) shows $V_1(x,y)$, the FT of the original image, while fig. 18(e) shows $V_{1H}(x,y)$, the FT of the modified image. However, in order to reconstruct the corresponding images, we require the complex FT's. These are illustrated next. Figures 18(b) and (c) show the real and imaginary parts respectively of $\tilde{U}_1(k_{x'}, k_{y'}) = \tilde{U}_1(x, y)$, the FT of the original image. Inverse FT of $\tilde{U}_1(x, y)$ yields the original image, $U_1(x',y')$, which is shown in axonometric view in fig 18(d). Note that the original image is real only. Similarly fig. 18 (f) and (g) show the real and imaginary parts, respectively, of $\tilde{U}_{1H}(k_{x'}, k_{y'}) = \tilde{U}_{1H}(x, y)$, the FT of the modified image. Inverse FT of $\tilde{U}_{1H}(x, y)$ yields the modified image, $U_{1H}(x',y')$. Note that $U_{1H}(x',y')$ is complex, as a result of the hole-cutting process. Real and imaginary parts of $U_{1H}(x',y')$ are shown in figure 18(h) and (i) respectively.

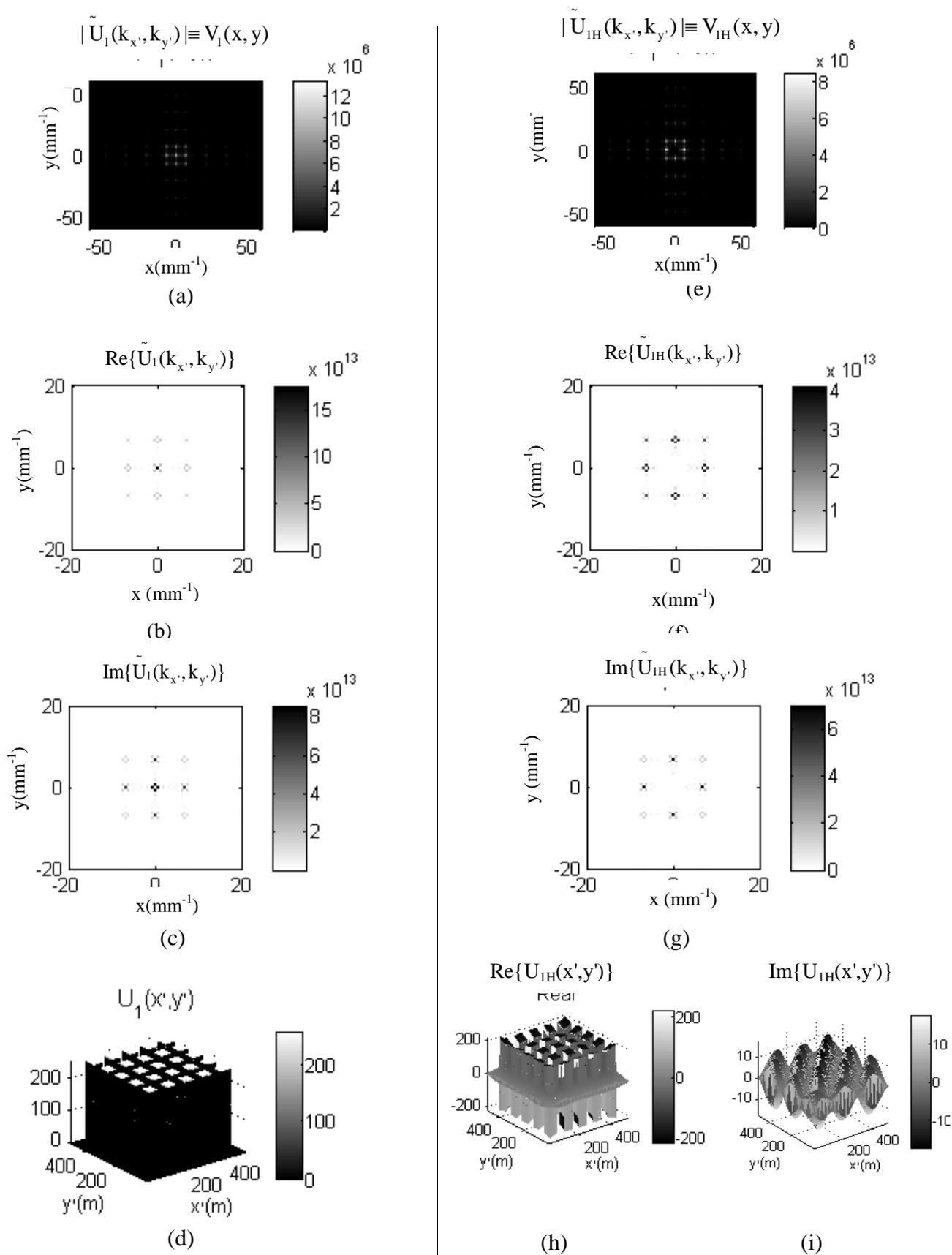

Fig 18: Illustration of the fact that cutting a hole of certain radius in the center of the FT of the image does not change the image significantly. See text for details. [Note that, in fig 16(b), (c), (f) and (g), the color has been inverted for clear visualization]

In fig. 19(a), we consider two images, $U_1(x',y')$ and $U_3(x',y')$, where $U_3(x',y')$ is scaled down by a linear factor of 2 and also rotated with respect to $U_1(x',y')$ by an angle of $\theta_0 = 30^0$. The magnitude of the FT of $U_3(x',y')$, denoted as $V_3(x,y)$, is also rotated by an angle of $\theta_0 = 30^0$ and the area is enlarged by a factor of 4, as shown in fig 19(b). In fig 19(c), a hole of radius $r_0$ is created in the center of each of $V_1(x,y)$ and $V_3(x,y)$, producing functions denoted as $V_{1H}(x,y)$ and $V_{3H}(x,y)$ respectively. The corresponding polar distribution, $F_1(r,\theta)$ and $F_3(r,\theta)$, are shown in fig 19(d). As shown in fig 19(e), the polar-logarithmic distributions are still identical in shape, except that $G_3(\rho,\theta)$, is shifted from $G_1(\rho,\theta)$ in both $\theta$-direction and $\rho$-direction. In the process described above, two original images, $U_1(x',y')$ and $U_3(x',y')$, are converted to PMT images, $G_1(\rho,\theta)$ and $G_3(\rho,\theta)$, which act as inputs to the HOC architecture. Fig 20(a) shows the final results of the HOC architecture where the cross correlation signal $T_3$ has two peaks at positions ($\rho' = \log(\sigma)$, $\theta' = \theta_0$) and ($\rho' = \log(\sigma)$, $\theta' = -(2\pi-\theta_0)$) and the anti cross-correlation signal, $T_4$ also has two peaks at positions ($\rho' = -\log(\sigma)$, $\theta' = -\theta_0$) and ($\rho' = -\log(\sigma)$, $\theta' = (2\pi-\theta_0)$). As mentioned in section 3, the peaks at positions ($\rho' = \log(\sigma)$, $\theta' = -(2\pi-\theta_0)$) and ($\rho' = -\log(\sigma)$, $\theta' = (2\pi-\theta_0)$) are very small compared to other peaks so that they are barely visible in the final output signal. Fig 20(b) shows that after thresholding the peaks of the cross-correlation signals at positions ($\rho' = \log(\sigma)$, $\theta' = \theta_0$) and ($\rho' = -\log(\sigma)$, $\theta' = -\theta_0$) are clearly visible. From the location of the peaks, we can infer that the objects are rotated with respect to each other by an angle of $30^0$ and also scaled down by a factor of 2.

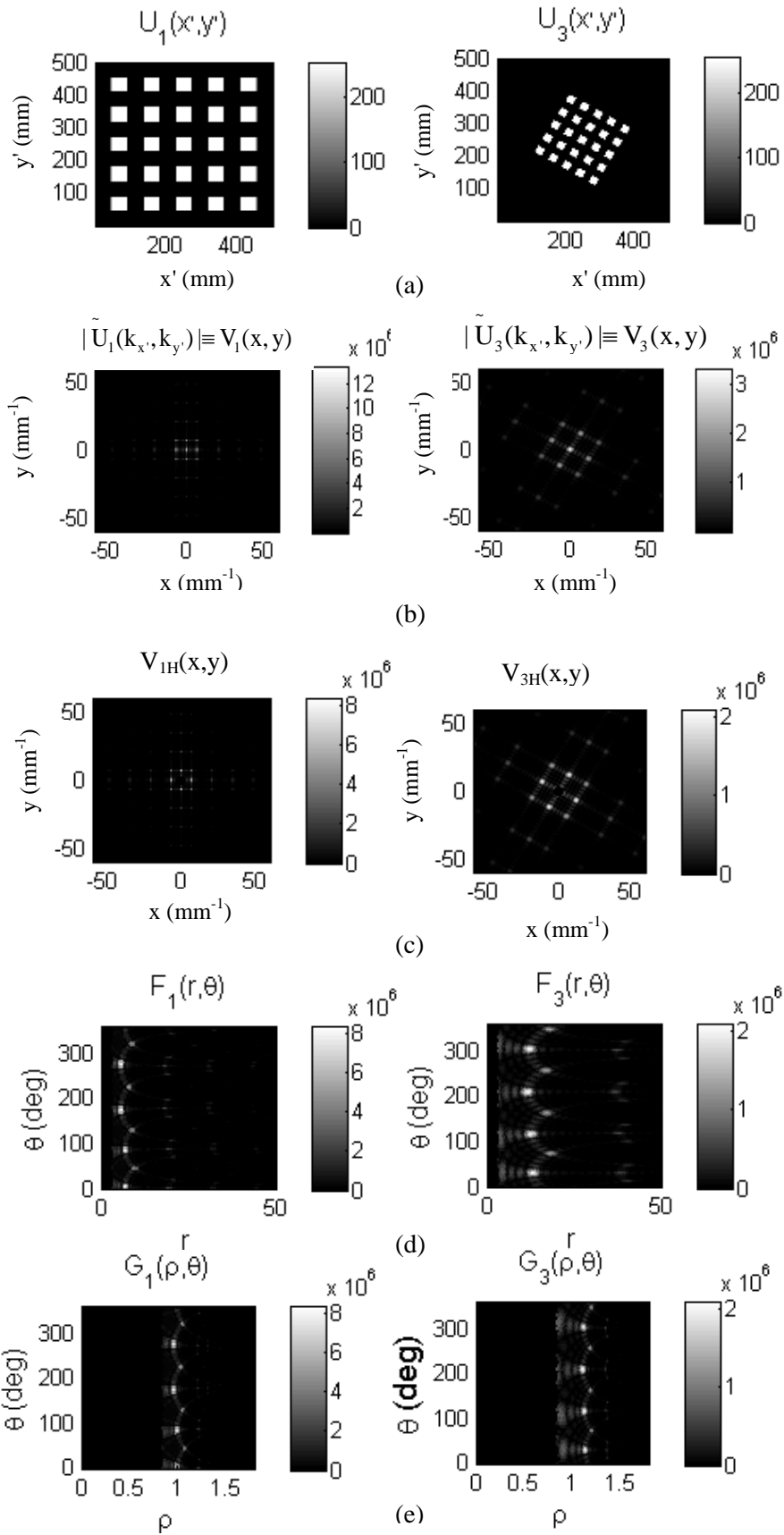

Fig. 19:(a) we consider two images $U_1(x',y')$ and $U_3(x',y')$, where $U_3(x',y')$ is smaller in area than $U_1(x',y')$ by a factor of 4 and also rotated by an angle of $\theta_0 = 30^0$ ; (b) The corresponding FTs are denoted as $V_1(x,y)$ and $V_3(x,y)$, which are similar in shape except $V_2(x,y)$ has a larger area than that of $V_1(x,y)$ by a factor of 4 and also rotated by an angle of $\theta_0 = 30^0$; (c) A hole of radius $r_0$ is created in the center of $V_1(x,y)$ and $V_3(x,y)$ which are denoted as $V_{1H}(x,y)$ and $V_{3H}(x,y)$ respectively; (d) The corresponding polar distributions, $F_1(r,\theta)$ and $F_3(r,\theta)$, are shifted in the $\theta$-direction by an amount of $\theta_0 = 30^0$ and also shifted in the r-direction by a linear scaling factor of 2. (e) The polar-logarithmic distributions, $G_1(\rho,\theta)$ and $G_3(\rho,\theta)$ are identical in shapes, except for a shift in the $\rho$-direction [equaling the logarithm of the scale factor: $\log(2) \cong 0.3$] and also a shift in the $\theta$-direction by an amount of $\theta_0 = 30^0$ .

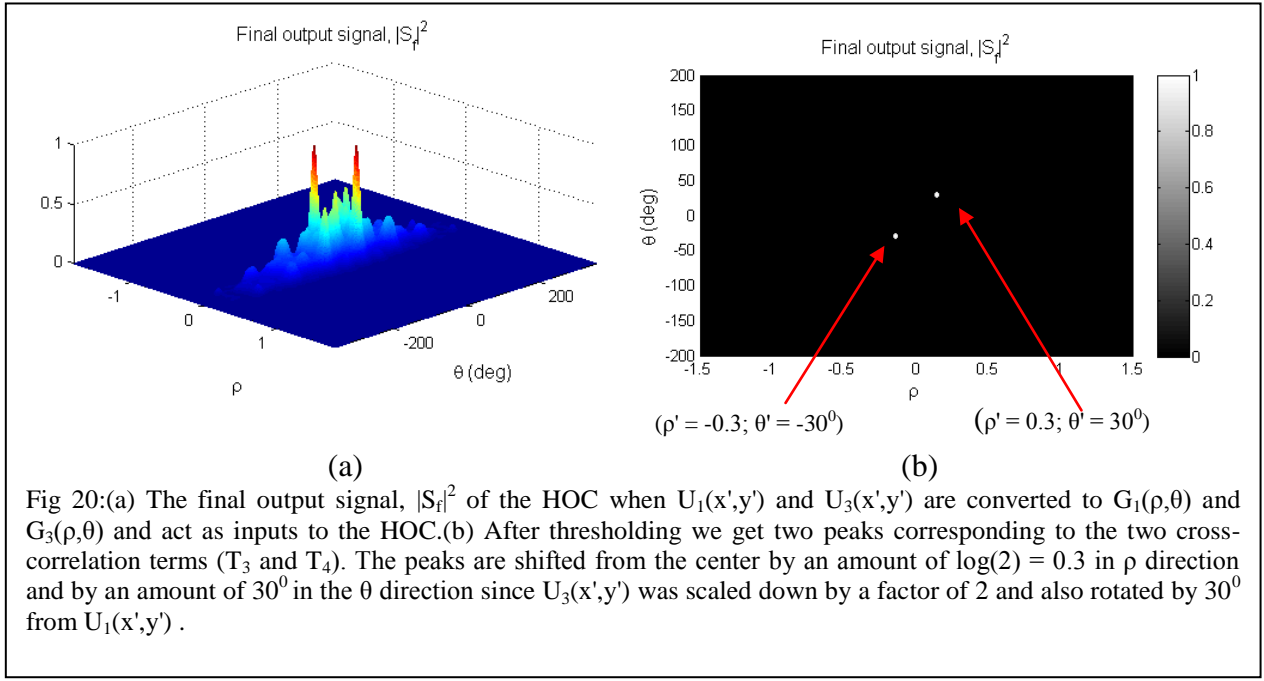

Fig 20:(a) The final output signal, $|S_f|^2$ of the HOC when $U_1(x',y')$ and $U_3(x',y')$ are converted to $G_1(\rho,\theta)$ and $G_3(\rho,\theta)$ and act as inputs to the HOC.(b) After thresholding we get two peaks corresponding to the two cross-correlation terms ($T_3$ and $T_4$). The peaks are shifted from the center by an amount of $\log(2) = 0.3$ in $\rho$ direction and by an amount of $30^0$ in the $\theta$ direction since $U_3(x',y')$ was scaled down by a factor of 2 and also rotated by $30^0$ from $U_1(x',y')$.

## 5. Multiple object detection using the HOC Architecture

In reference 7, we showed how to recognize a single object using the HOC architecture in a shift invariant manner. In this paper so far, we have shown how to recognize a single object in a shift, scale and rotation invariant manner. However, there are potential scenarios where the query field may contain multiple matches to the reference object. In this section, we describe how to achieve distinct detection of these multiple matches. We consider two different scenarios.

First, we consider the case where the multiple images in the query field are only shifted, without any scale change and rotation angle. In this case, the architecture needed does not employ the PMT process. However, it is somewhat different from the approach used for detecting multiple (unscaled and unrotated) matches using a conventional holographic correlator, because of the fact that the HOC produces both cross-correlation and anti-crosscorrelation signal. Second, we consider the case where the multiple images in the query field have potentially distinct values of shift, scale factor and rotation angle. In this case, the PMT process has to be employed. However, since the PMT process eliminates the shift information, multiple object recognition in this case requires a substantially different architecture.

*5.1 Multiple objects detection for shifted images without rotation and scale change:*

Detection of multiple objects using the HOC architecture without rotation and scale change is similar to single object detection under the same scenario (i.e., without rotation and scale change), in that it does not require using the PMT process. However, a potential complication arises due to the presence of both $T_3$ (cross-correlation) and $T_4$ (anti-cross correlation) terms. Specifically, in the presence of multiple matches, it becomes difficult to determine whether a peak corresponds to $T_3$ of a given matched image, or $T_4$ of another matched image, for example. This can be circumvented by applying the following technique. Assume first that the reference image is represented by a grid of NXN points. In contrast, we confine the query image, potentially containing multiple objects, to a grid of only N/2 x N points. We now map this query image to a grid of NXN points, thus producing a final query image where half the image is blank.

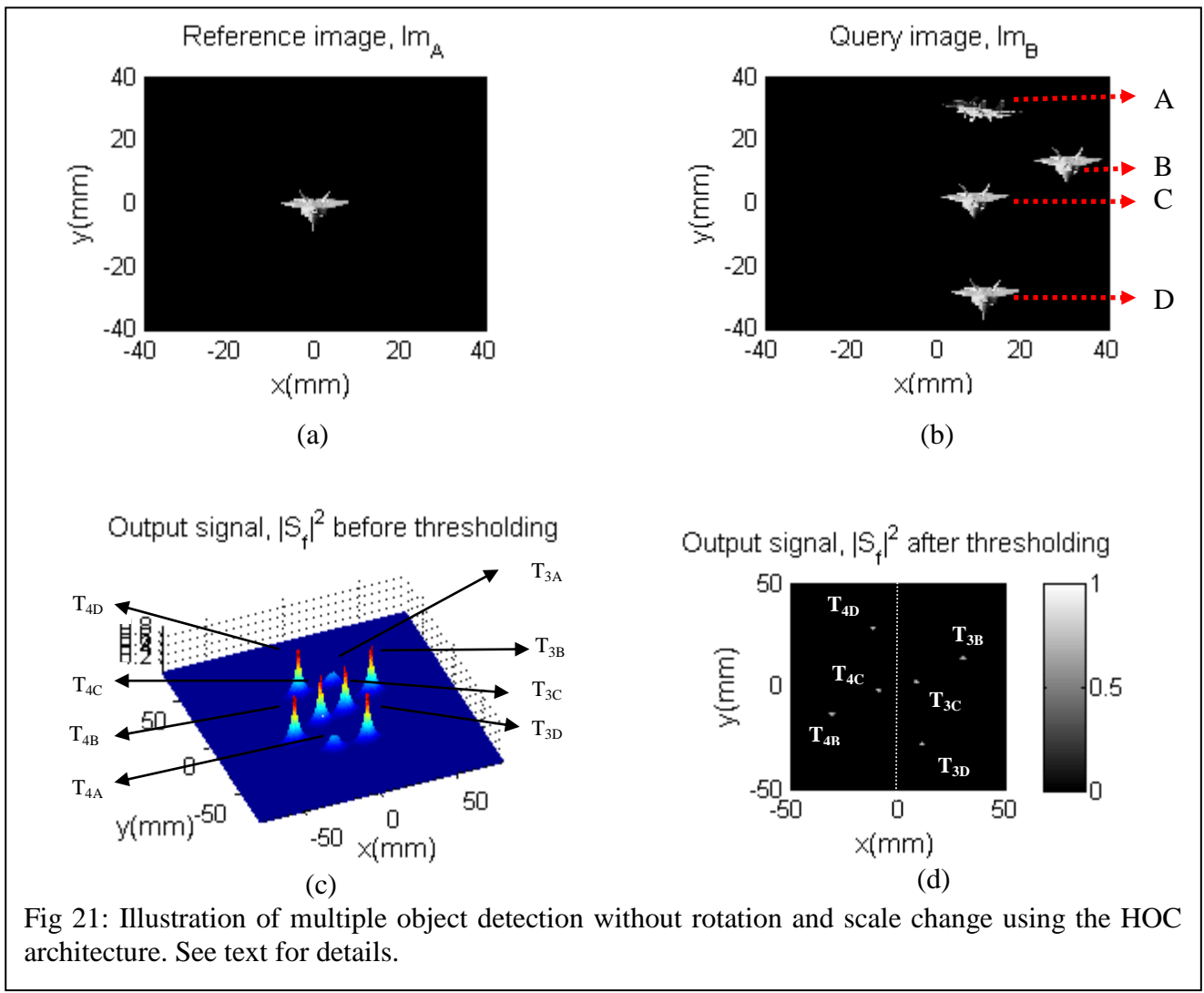

Fig 21: Illustration of multiple object detection without rotation and scale change using the HOC architecture. See text for details.

Consider a situation where the blank half of the query image is on the left side. It is easy to see that, after carrying out the correlation process in the manner described in section 2, the peaks representing all the $T_3$'s (corresponding to multiple matches) will appear on the right side of the final signal plane, while all the $T_4$'s will appear on the left side of the final signal plane, thus avoiding the potential ambiguities between the $T_3$'s and the $T_4$'s mentioned above.

Fig. 21 illustrates the results of a simulation for multiple object detection with the HOC using this approach. Fig. 21(a) shows the reference image, $Im_A$, plotted on an N X N grid. Fig 21(b) show the query image, $Im_B$, which contains three images that match the reference (denoted as B,C and D), and one that does not (denoted as A). Note that these four images are confined to the right half plane only, leaving the left half blank. Fig 21(c) shows the output signal, $|S_f|^2$ where we can see six sharp peaks corresponding to cross-correlation signals ($T_{3B}$, $T_{3C}$ and $T_{3D}$) and anti-cross correlation signals ($T_{4B}$, $T_{4C}$ and $T_{4D}$) of the three matched images. The other two signals in the output plane with lower peaks correspond to the cross-correlation ($T_{3A}$) and anti cross-correlation ($T_{4A}$) signals for the unmatched case. After thresholding (as shown in Fig 21(d)), the final signals are clearly visible from which we can infer that three matches are found. In addition, the distances of the peaks from the center reveal the locations of the matched images.

This approach of leaving a blank space also has an added advantage, in that it ensures that the overlap parameter, η (defined in section 2) is never less than unity. This can be seen clearly by considering image C, which is located at the left edge of the right half plane. Since it is contained fully in the right half plane, the distance between this and the reference images is $|\overrightarrow{\rho_m}|$, which corresponds to η = 1. All other images that are further away from the boundary between the left and right half planes would thus have a value of η > 1. Therefore, the cross-correlation and anti-cross correlation signals will be clearly revolved for all images.

Finally, we note that the use of the rectangular field (N/2 X N) in confining the query image may be inconvenient in some situations, especially if the camera used in acquiring the query image has an image field that is a square. This can be circumvented by confining the query image to a square field with N/2 x N/2 points, leaving the other three quadrants blank.

*5.2 Multiple objects detection in shift, scale and rotation invariant manner using the HOC architecture:*

Next, we consider a situation where the query field contains multiple replicas of the reference image, but each with potentially a different position, a different scale factor, and a different angular orientation. Obviously, this case would require the use of the PMT process. However, the PMT process loses the information about the relative position of any image once the phase information in the FT is eliminated by measuring the magnitude of the FT. Thus, the magnitude of the FT's of each of the matched images in the query field will overlap with one another, making it impossible to find any matches.

To overcome this problem, we propose the approach illustrated schematically in fig. 22 and 23. The situation of interest here is as follows. We assume that we have one reference image, and $L \equiv n/2 \times n$ captured query images (n is an integer), each of which is a potentially shifted, scaled and rotated replica of the reference image. In principle, we can employ the PMT enhanced HOC process L times. The goal here is to carry out these L correlations simultaneously.

To start, we fit each captured image into a grid of $(N/n) \times (N/n)$, where $N \times N$ is the grid size for the reference image. We then fit these images into the right half plane of an $N \times N$ grid. This is illustrated in fig. 22. Next, we use an SLM to convert the query image to the optical domain. However, instead of sending the whole query image at once, we send to the SLM only one of the small grids (of size N/n x N/n) on the right half plane with all the other small grids being dark. This is illustrated schematically on the SLM screen shown on the left edge of the fig. 23, for the first row and n/2-th column of the right half plane, for example. The image in this grid is denoted as $Q_{1,n/2}$. The lens, the FPA and the FPGA, as shown in the rest of fig 23, are used to produce the corresponding PMT image, denoted as $G_{1,n/2}(\rho,\theta)$. This process is repeated $n^2/2$ times, which is the number of small grids containing images, without changing the positions of the SLM, the lens and the FPA. The $n^2/2$ numbers of PMT images produced and stored in the FPGA are then mapped to a corresponding set of small grids, which in turn is sent to the final SLM (SLM-2) which in turn is sent to the final SLM for detecting cross-correlation and anti-cross correlation signals.

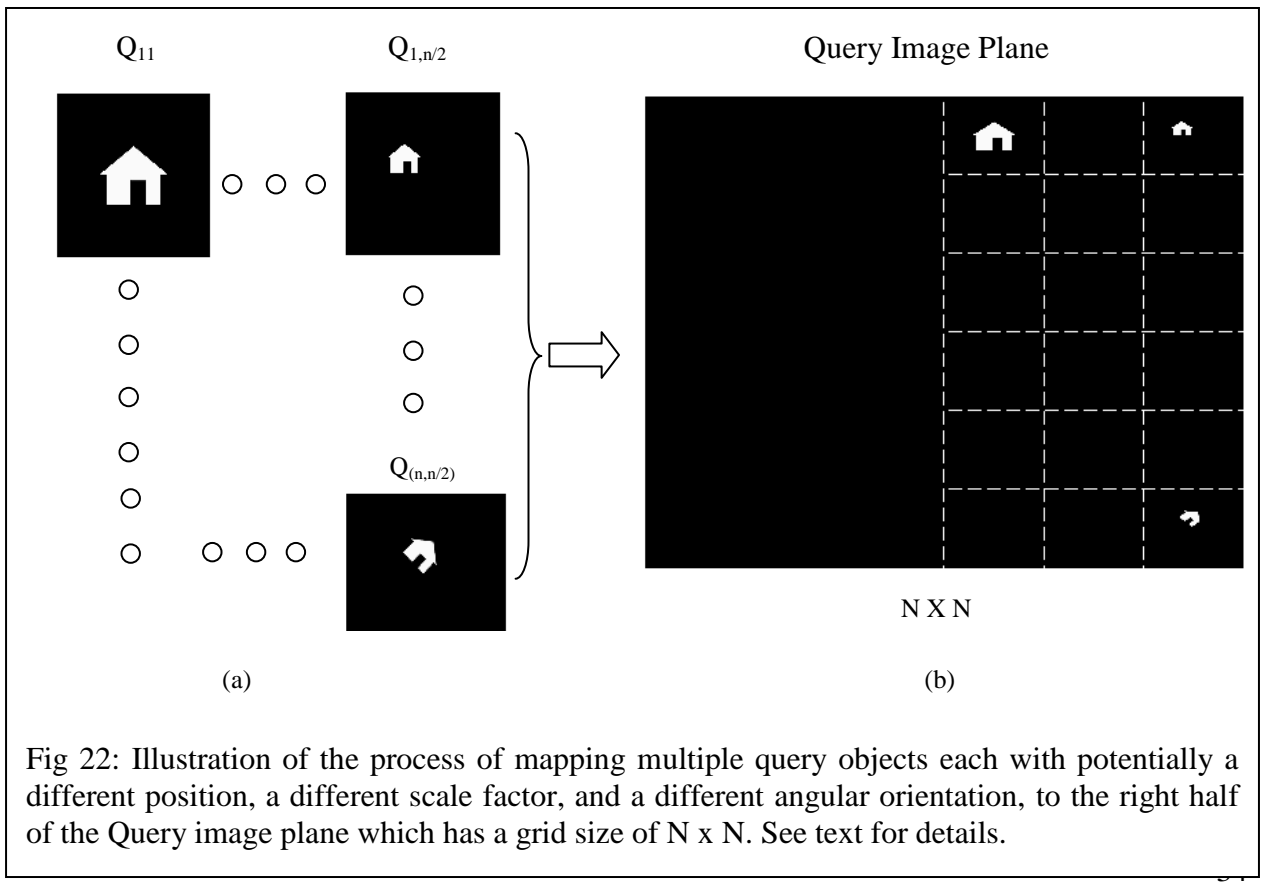

Fig 22: Illustration of the process of mapping multiple query objects each with potentially a different position, a different scale factor, and a different angular orientation, to the right half of the Query image plane which has a grid size of N x N. See text for details.

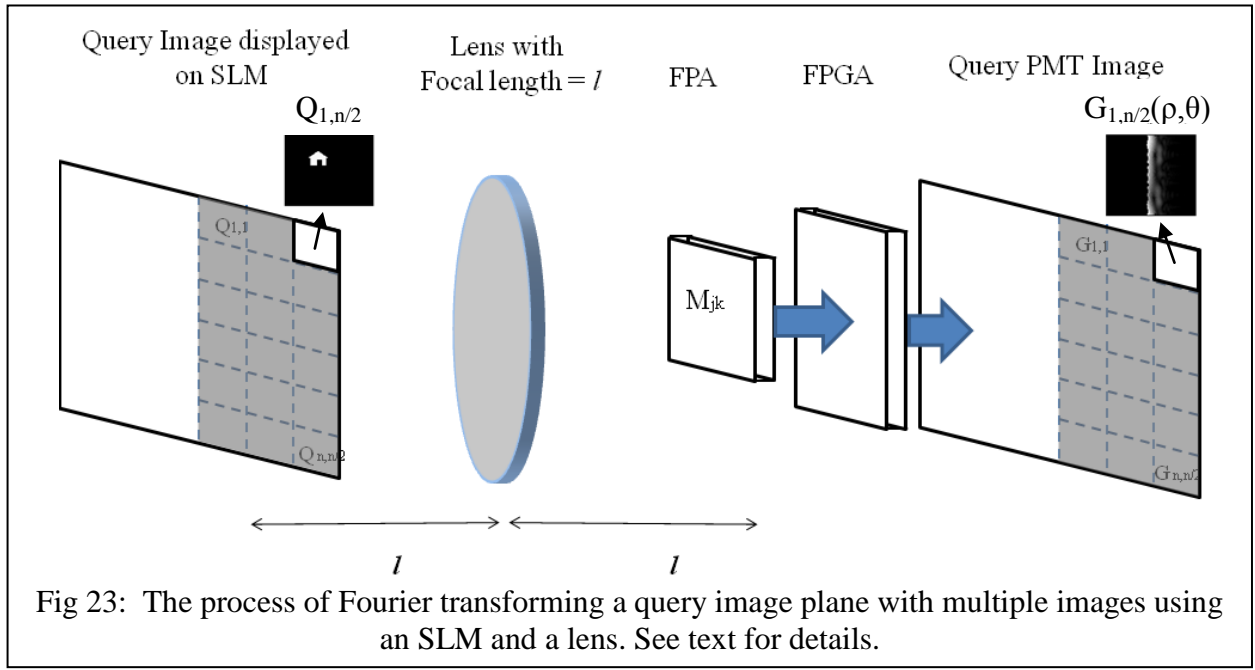

Fig 23: The process of Fourier transforming a query image plane with multiple images using an SLM and a lens. See text for details.

In fig 24, we show results of numerical simulations used to illustrate the process described above. For simplicity, we have used an artificial reference image that has a clear hole in its FT's, similar to that shown earlier in fig 11. The FT of the reference image is shown in fig 24(a), denoted as $F_{00}$, and the FTs of multiple query objects are shown in fig 24(b), which are denoted as $F_{ij}$ (i = row number; j = column number). For example, $F_{42}$ is the image shown on the bottom right corner. The left half of the query plane is kept blank for the reason described above. As can be seen in fig 24(b), $F_{21}$ and $F_{42}$ are similar to $F_{00}$; $F_{22}$ and $F_{31}$ are rotated from $F_{00}$ by an angle of $\theta_0 = 30^0$ ; $F_{11}$ and $F_{32}$ are scaled down from $F_{00}$ by a factor of $\sigma = 2$; $F_{12}$ and $F_{41}$ are scaled down by a factor of $\sigma = 2$ and also rotated by an angle of $\theta_0 = 30^0$ from $F_{00}$. The corresponding PMT images, $G_1(\rho,\theta)$ and $G_2(\rho,\theta)$ are shown in fig 24(c) and 24(d), respectively [16]. Fig 24(e) shows the final signal $|S_f|^2$ after thresholding. The right half side shows the cross-correlation signals and the left side shows the anti-cross- correlation signals. The red dots in fig 24(e) correspond to the auto-correlation of the reference PMT image and the white dots on the right half plane represent the corresponding cross-correlation signals. From the distance between the red dot and the white dot in a given box, we can infer the scale and rotation change between the reference image and the query image.

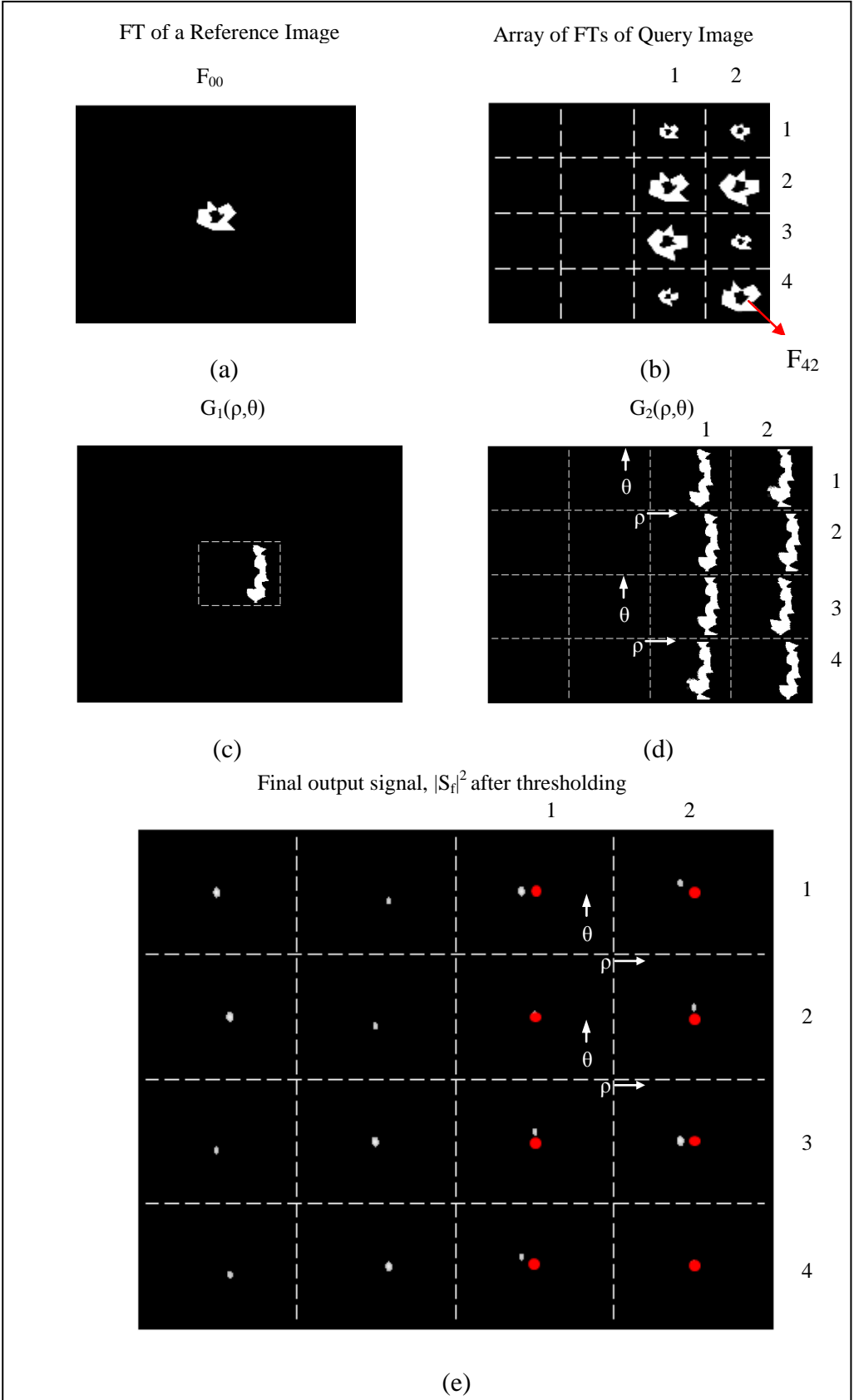

Fig 24: (a) FT of the reference image (b) The array of FTs of multiple objects where one side is intentionally left blank. (c) PMT of the reference image (d) PMT of the array of query images (e) Final results from the HOC. We have to consider only the right half of this plane. See text for details.

## 6. Conclusion

We have shown that our proposed hybrid optoelectronic correlator (HOC), which correlates images using spatial light modulators (SLM), detectors and field programmable gate array (FPGA), is capable of detecting objects in a scale and rotation invariant manner, along with the shift invariance feature, by incorporating polar mellin transform (PMT). We have also illustrated a key limitation of the ideas presented in previous papers on performing PMT and presented a solution to circumvent this limitation by cutting out a small circle at the center of the Fourier Transform which precedes PMT. Furthermore, we showed how to carry out shift, rotation and scale invariant detection of multiple matching objects simultaneously, a process previously thought to be incompatible with PMT based correlators. We presented results of numerical simulations to validate the concepts. Experimental efforts are underway in our laboratory to demonstrate these capabilities of the HOC using bulk components. Efforts are also underway to develop an integrated graphic processing unit (IGPU) [7] in order to realize a high-speed version of the HOC.

## 7. Acknowledgement

This work is supported by AFOSR Grant FA9550-10-01-0228, NSF CREST award 1242067 and NASA URC Group-V award NNX09AU90A.

16. Recall that a PMT image is plotted as a function of $\rho$ and $\theta$. In the $\theta$ direction, any image will cover the whole range from 0 to $2\pi$. As a result, two PMT images in two boxes adjacent in the vertical direction will tend to merge into each other. This problem is circumvented by scaling each PMT image to 90% of its actual size, thus creating a guard band. This step does not affect the outcome of the correlator process.